\documentclass[
	aps,prd,
	reprint,
	showkeys,
	amsmath,amssymb,
	nofootinbib,
	floatfix,
        superscriptaddress,
	]{revtex4-2}
\usepackage[utf8]{inputenc}
\usepackage{aas_macros}
\pdfoutput=1
\usepackage{hyperref}

\usepackage[caption=false]{subfig}
\usepackage{graphicx}
\hypersetup{ 
			colorlinks=true,
			pdfauthor={Georgios Valogiannis and Cora Dvorkin},
			pdftitle={WST TBD},
			citecolor=blue
			}

\usepackage{xcolor}
\usepackage{multirow}
\usepackage{array} 
\usepackage{ctable}

\newcommand{\ahod}{\textsc{AbacusHOD}}

\begin{document}

\title{Precise Cosmological Constraints from BOSS Galaxy Clustering with a Simulation-Based Emulator of the Wavelet Scattering Transform}

\author{Georgios Valogiannis}
 \email{gvalogiannis@g.harvard.edu}
 \affiliation{
 Department of Physics, Harvard University, Cambridge, MA, 02138, USA\\
}
 \affiliation{
 Department of Astronomy and Astrophysics, University of Chicago, Chicago, IL, 60637, USA\\
}
 \author{Sihan Yuan}
\email{sihany@stanford.edu}
 \affiliation{
Kavli Institute for Particle Astrophysics and Cosmology, 452 Lomita Mall, Stanford University, Stanford, CA 94305,USA\\
}
 \affiliation{
SLAC National Accelerator Laboratory, 2575 Sand Hill Road, Menlo Park, CA 94025, USA\\
}
\author{Cora Dvorkin}
 \email{cdvorkin@g.harvard.edu}
\affiliation{
 Department of Physics, Harvard University, Cambridge, MA, 02138, USA\\
}

\begin{abstract}
We perform a reanalysis of the BOSS CMASS DR12 galaxy dataset using a simulation-based emulator for the Wavelet Scattering Transform (WST) coefficients. Moving beyond our previous works, which laid the foundation for the first galaxy clustering application of this estimator, we construct a neural net-based emulator for the cosmological dependence of the WST coefficients and the 2-point correlation function multipoles, trained from the state-of-the-art suite of \textsc{AbacusSummit} simulations combined with a flexible Halo Occupation Distribution (HOD) galaxy model. In order to confirm the accuracy of our pipeline, we subject it to a series of thorough internal and external mock parameter recovery tests, before applying it to reanalyze the CMASS observations in the redshift range $0.46<z<0.57$. We find that a joint WST + 2-point correlation function likelihood analysis allows us to obtain marginalized 1$\sigma$ errors on the $\Lambda$CDM parameters that are tighter by a factor of $2.5-6$, compared to the 2-point correlation function, and by a factor of $1.4-2.5$ compared to the WST-only results. This corresponds to a competitive $0.9\%$, $2.3\%$ and $1\%$ level of determination for parameters $\omega_c$, $\sigma_8$ $\&$ $n_s$, respectively, and also to a $0.7\%$ $\&$ $2.5 \%$ constraint on derived parameters h and $f(z)\sigma_8(z)$, in agreement with the \textit{Planck} 2018 results. Our results reaffirm the constraining power of the WST and highlight the exciting prospect of employing higher-order statistics in order to fully exploit the power of upcoming Stage-IV spectroscopic observations. 
\end{abstract}

\maketitle

%%%%%%%%%%%%%%%%%%%%%%%%%%%%%%%%%%%%%%%%%%%%%%
\section{Introduction \label{sec:intro}}

The advent of precision cosmology, with a large collection of surveys including the Dark Energy Spectroscopic Instrument (DESI) \citep{Levi:2013gra,2016DESI}, the {\it Vera C. Rubin} Observatory Legacy Survey of Space and Time (LSST) \citep{Abell:2009aa,Abate:2012za}, {\it Euclid} \citep{Laureijs:2011gra}, and the {\it Nancy Grace Roman} Space Telescope \citep{Spergel:2013tha}, that will accurately probe the 3-dimensional (3D) large-scale structure (LSS) of the universe, promises to dramatically change our fundamental understanding of the cosmos. Among the wealth of valuable information offered by cosmological observations of this kind, lies the opportunity to tackle major open questions in modern physics, such as the source of the accelerated expansion of the universe at late times \citep{Copeland:2006wr}, the nature of dark matter \citep{LSSTDarkMatterGroup:2019mwo}, the large-scale properties of gravity \citep{Ishak:2018his,doi:10.1146/annurev-astro-091918-104423, Alam:2020jdv}, the properties of massive neutrinos \citep{LESGOURGUES2006307,Dvorkin:2019jgs}, as well as the physics of the primordial universe and other light relics \citep{Chen_2016,DePorzio:2020wcz, Xu:2021rwg}.

The probability distribution that describes the observed large-scale structure of the universe at late times is known to deviate from the familiar Gaussian form characterizing the primordial density field. The nonlinear process of gravitational instability, responsible for the formation of the 3D cosmic web, imparts a non-Gaussian distribution in the observed large-scale structure of the universe. As a consequence, the standard compression achieved by the 2-point correlation function of density fluctuations fails to capture all available information encoded in the clustered field \citep{PhysRevLett.108.071301}. Even though working with the 2-point function statistics has sufficed in traditional applications of cosmological parameter inference up until recently, such an approach will be inadequate if the potential of the upcoming generation of cosmological surveys is to be fully exploited. Accurately modeling structure formation down to the nonlinear regime in principle requires the inclusion of higher-order moments as a part of the traditional parameter inference, a line of research that is currently very actively pursued \citep{10.1093/mnras/stv961,10.1093/mnras/stw2679,BERNARDEAU20021,Hahn_2020, Hahn_2021,PhysRevD.105.043517,Chen:2021vba,Philcox:2021hbm,2023arXiv231015243H}. Nevertheless, the requirements associated with handling $n$-point correlation functions, both in terms of the necessary computational cost of evaluation, but also due to the relatively large dimensionality of the final data vector, quickly render such an approach intractable when going to higher $n$. Even when these challenges can be tempered using different kinds of techniques, the total information encoded in a non-Guassian field has been shown to escape the entire correlation hierarchy, with the magnitude of loss getting progressively more pronounced with increasing degree of non-Gaussianity \citep{PhysRevLett.108.071301}. 

The obstacles mentioned above motivate developing novel ways of accessing the additional information that lies beyond the linear regime, using summary statistics that are sensitive to higher-order information, but yet impose minimal additional computational burden compared to a standard power spectrum evaluation. Among the long list of estimators of this kind that have been considered in the literature \footnote{For a recent exploration of higher-order statistics in the particular context of weak lensing, also see Ref. \citep{2023arXiv230112890E}.}, this active subfield involves proxy estimators \citep{PhysRevD.91.043530,Dizgah_2020,Chakraborty:2022aok, 2023JCAP...03..045H,2024arXiv240113036C,2024arXiv240115074H}, efforts to isolate the information encoded in the cosmic voids of the LSS \citep{Pisani:2019cvo,Massara_2015, 10.1093/mnras/stz1944, 10.1093/mnras/stv777, Hamaus_2015, Kreisch:2021xzq,Bonnaire:2021sie, 2023arXiv230205302R}, nonlinear transformations that partly restore the Gaussianity of the density field \citep{Neyrinck_2009,PhysRevLett.108.071301,PhysRevLett.107.271301,White_2016,PhysRevD.97.023535,PhysRevLett.126.011301,2022arXiv220601709M}, splitting the density field into different environments \citep{10.1093/mnras/stab1654,2022arXiv220904310P,Bayer:2021iyb,Paillas2023:2309.16541,Cuesta-Lazaro2023:2309.16539}, working with k-nearest neighbors \citep{Banerjee:2020umh,10.1093/mnras/stab961} and a variety of other beyond 2-point statistics, such as Minkowski functionals \citep{10.1046/j.1365-8711.1999.02912.x,10.1111/j.1365-2966.2012.21103.x,10.1093/mnras/stt1316,PhysRevLett.118.181301,2023arXiv230208162L}, the minimum spanning tree \citep{Naidoo:2021dxz} or 1-point statistics \citep{Uhlemann:2019gni,2020PhRvD.102l3546J}. The recent rapid evolution of Artificial Intelligence (AI) has motivated efforts to extract cosmological non-Gaussianities using Convolutional Neural Networks (CNNs) \citep{6522407}, demonstrating great promise in idealized settings \citep{PhysRevD.97.103515,Villaescusa_Navarro_2021, Perez:2022nlv,Dvorkin:2022pwo}. Whether and how this simulation-based performance can be extended to reliable interpretations of actual galaxy data is still a matter of study; for example see Ref. \citep{Hahn:2023udg}.

Another path towards harnessing the nonlinear information encoded in the LSS, can be carved by seeking for a balanced trade-off between performance and interpretability, working in the middle-ground between traditional clustering estimators and CNNs. Such a trade-off is attempted by the Wavelet Scattering Transform (WST), \citep{https://doi.org/10.1002/cpa.21413,6522407}, which was first proposed in the context of computer vision. In direct analogy to the architecture of a CNN, a scattering network is constructed by successively performing two operations to an input field: wavelet convolution and modulus. After averaging over all pixels, the resulting outcome is a basis of interpretable WST coefficients, which can quantify the clustering information in the input field \citep{Sifre_2013_CVPR,10.1214/14-AOS1276,6822556}, while avoiding the previously discussed limitations of the standard moment expansion \citep{PhysRevLett.108.071301}. Motivated by these attractive properties, the WST has recently seen successful applications across the spectrum of natural sciences \citep{cheng:2021xdw}, including astrophysics \citep{refId0, Saydjari_2021,refdust}, cosmology \citep{10.1093/mnras/staa3165,10.1093/mnras/stab2102,PhysRevD.105.103534,PhysRevD.106.103509,PhysRevD.102.103506,10.1093/mnras/stac977,2022arXiv220407646E,2023arXiv231015250R,10.1093/mnras/stac2662,10.1093/mnras/stac977} and molecular chemistry \citep{10.5555/3295222.3295400,doi:10.1063/1.5023798}. 

As far as 3D clustering explorations are concerned, the first WST application was performed by Ref. \citep{PhysRevD.105.103534}, working with the fractional matter overdensity field obtained by N-body simulations \citep{Villaescusa_Navarro_2020} as input. Through a Fisher forecast, the basis of WST coefficients up to $2^{nd}$ order was found to predict a substantial improvement on the 1-$\sigma$ errors obtained on 6 cosmological parameters, exceeding the performance of both the standard and also the marked power spectrum. Another application was subsequently performed by Ref. \citep{2022arXiv220407646E}, finding similar levels of improvement. Building upon these encouraging results, the subsequent work of Ref. \citep{PhysRevD.106.103509} developed the first application of the WST to actual galaxy data, analyzing observations from  the CMASS sample of the Baryon Oscillation Spectroscopic Survey (BOSS) \citep{Eisenstein_2011,Dawson_2012} (under some approximations, however, as we explain in the next paragraph). The WST, once again, was found to deliver a notable improvement to the errors obtained on 4 cosmological parameters, which were 3-6 times tighter compared to the ones from the galaxy power spectrum. This analysis demonstrated the great promise held in the use of the WST as a means of parameter inference in the context of spectroscopic surveys and precision cosmology in general. 

Even though Ref. \citep{PhysRevD.106.103509} laid out all the necessary steps to account for the complexities related to a WST application to spectroscopic galaxy data, combined with a set of high-fidelity galaxy mocks, it adopted a Taylor expansion approximation to model the cosmological dependence of the WST coefficients, which in principle could fail to capture non-Gaussianities present in the parameter likelihood. As a result, the accuracy of this approach was not tested in recovery tests against other simulations. In this work, we move beyond these approximations, and revisit our previous analysis with a full emulator predicting the cosmological dependence of the WST estimator. We take advantage of the full extent of the state-of-the-art suite of \textsc{AbacusSummit} simulations \citep{2021Maksimova}, which consists of a broad grid exploring variations in 8 cosmological parameters, in combination with a semi-analytic model to parametrize the physics of galaxy formation for each cosmology. This extended suite enables the training of a neural net-based emulator that predicts the cosmological dependence of the WST coefficients in a 15-dimensional parameter space. In order to quantify the accuracy of this emulator, we subject it to a series of thorough parameter recovery tests against hold-out simulations, as well as against simulations using different models to capture small-scale galaxy physics. After we confirm that our model satisfies the necessary levels of accuracy for a reliable cosmological application, we use it to reanalyze the BOSS CMASS galaxy dataset, and obtain the marginalized $1-\sigma$ errors on 4 $\Lambda$CDM cosmological parameters, as well as on extended scenarios. We contrast our results against the ones obtained by the standard analysis performed using the multipoles of the anisotropic correlation function of galaxies, and discuss how our analysis compares to our previous work and prior ones in the literature. 

Our paper is structured as follows: in \S\ref{WST:sup} we introduce the Wavelet Scattering Transform and in \S\ref{sec:Dataset} we describe the BOSS dataset. We then proceed to lay out all the ingredients used to construct our simulation-based forward model in \S\ref{WST:model}, as well as the details of our analysis pipeline in \S\ref{sec:Analysis}. Finally, we present our results in \S\ref{sec:Results}, before concluding in \S\ref{sec:Conclusions}. More technical details are discussed in Appendices \S\ref{Appsec:Gaussianity}, \S\ref{Appsec:HOD}, \S\ref{Appsec:Cemu}, and \S\ref{Appsec:omegabpriors}.

%%%%%%%%%%%%%%%%%%%%%%%%%%%%%%%%%%%%%%%%%%%%%%
\section{Wavelet Scattering Transform}\label{WST:sup}

The Wavelet Scattering Transform  \citep{https://doi.org/10.1002/cpa.21413,6522407} is a novel summary statistic that was proposed as an ideal middle-ground between a CNN and more traditional statistical estimators. Defined by a series of well-understood and interpretable mathematical operations, it can quantify the degree of clustering of an input field in a manner that not only matches, but also supersedes the properties of the standard 2-point correlation function \citep{https://doi.org/10.1002/cpa.21413}.

According to the WST definition, the two fundamental properties to which an input field, $I({\bf x})$, is subjected, are wavelet convolution and modulus. Specifically, given a localized wavelet probing a scale $j_1$ and an orientation $l_1$, that we will hereafter denote by $\psi_{j_1, l_1}(\bold{x})$, the WST transforms the input field as follows:
\begin{equation}\label{eq:convmod}
I'(\bold{x}) = |I(\bold{x}) \ast \psi_{j_1, l_1}(\bold{x}) |,   
\end{equation}
where $\ast$ indicates the convolution operation. If we further average over the transformed field in Eq. \eqref{eq:convmod}, we can derive a single number globally characterizing the field, which is called a WST coefficient. Furthermore, if the above sequence of elementary operations is successively repeated $n$ times, and for a range of different $j_1$ scales and $l_1$ angles covered by a family of localized wavelets, $\psi_{j_1, l_1}(\bold{x})$, it will form a $\textit{scattering network}$, with WST coefficients, $S_n$, given by: 

\begin{align}\label{eq:WSTcoeff:base}
 S_0 &= \langle |I(\bold{x})|\rangle, \nonumber \\
 S_1(j_1,l_1) &= \left\langle |I(\bold{x}) \ast \psi_{j_1, l_1}(\bold{x}) | \right\rangle, \\
 S_2(j_2,l_2,j_1,l_1) &= \left\langle |\left(|I(\bold{x}) \ast \psi_{j_1, l_1}(\bold{x}) |\right) \ast \psi_{j_2, l_2}(\bold{x}) |\right\rangle \nonumber,
\end{align}
explicitly shown up to order $n=2$ above. The angular brackets, $\langle . \rangle$, in Eq. \eqref{eq:WSTcoeff:base} and hereafter will denote taking the average value over the volume of the field\footnote{Formally defined as the expectation value.}. Convolving with a localized wavelet essentially quantifies the strength of clustering in the input field over the relevant scales, similar to the 2-point function. The WST coefficients of order $n$ have been shown to capture information related to the correlation function of order up to $2^n$ \citep{https://doi.org/10.1002/cpa.21413,10.1214/14-AOS1276}. Building upon this property, it follows that the hierarchy of Eqs. \eqref{eq:WSTcoeff:base} leads to a collection of WST coefficients that can quantify the higher-order clustering information of the input physical field, $I(\bold{x})$, in analogy to the moment expansion usually applied to cosmological density fields. Opposite to the conventional series of correlation functions, however, the WST has been found to be more efficient at extracting information out of an input field, especially in highly non-Gaussian cases which are particularly challenging for higher-order moments to accurately describe \citep{PhysRevLett.108.071301,cheng:2021xdw}. Furthermore, the fact that the input field always enters Eq. \eqref{eq:WSTcoeff:base} in a linear fashion guarantees a greater degree of numerical stability and robustness against outliers. In addition, the generated basis of WST coefficients is compact, such that the dimensionality of the resulting data vector can be kept under better control \citep{cheng:2021xdw}. It is worth noting that the operations of  wavelet (kernel) convolution, modulus (nonlinearity) and averaging (pooling), all implemented in a hierarchical scattering network, resemble the architecture and properties of a CNN with fixed kernels \citep{https://doi.org/10.1002/cpa.21413,6522407}. Combining all of the above properties, it becomes clear how the WST can be viewed as an interpretable alternative that lies between conventional summary statistics and CNNs, making it a potentially powerful tool to employ when harnessing higher-order information. In this work we will focus on the use of the WST for cosmological parameter inference, but we note that it can also be used in other applications, such as field synthesis and texture characterization, further discussed in Ref. \citep{cheng:2021xdw}.

Even though in the standard WST definition the input field enters Eq. \eqref{eq:WSTcoeff:base} linearly, slightly relaxing this assumption and allowing for $I(\bold{x})$ to be raised to a power $q$, instead, results in the following variant:
\begin{align} \label{eq:WSTcoeff:power}
 S_0 &= \langle |I(\bold{x})|^q\rangle, \nonumber \\
 S_1(j_1,l_1) &= \left\langle |I(\bold{x}) \ast \psi_{j_1, l_1}(\bold{x}) |^q \right\rangle, \\
 S_2(j_2,l_2,j_1,l_1) &= \left\langle |\left(|I(\bold{x}) \ast \psi_{j_1, l_1}(\bold{x}) |\right) \ast \psi_{j_2, l_2}(\bold{x}) |^q\right\rangle \nonumber,
\end{align}
which can lead to very interesting implications for cosmology, given that values of $q>1$ or $q<1$ respectively emphasize overdense or underdense regions of the LSS. This option was explored in the 3D matter overdensity WST application of Ref. \citep{PhysRevD.105.103534}, and was indeed found to produce more competitive constraints on cosmological parameters, with an emphasis on the sum of neutrino masses, when cosmic voids where highlighted using values of $q<1$. In this application we will stay aligned with our previous work \citep{PhysRevD.106.103509} and proceed with the version of WST given in Eq. \eqref{eq:WSTcoeff:power}.

Given that in this work we will focus on a WST application to 3D galaxy clustering, as we will specify below, the input field $I(\bold{x})$ will be taken to be 3-dimensional, even though the above discussion can in principle be valid for an arbitrary number of dimensions. Following our previous works \citep{PhysRevD.105.103534,PhysRevD.106.103509}, we adopt a mother wavelet given by the solid harmonic expression of
\begin{equation}\label{solid:sup}
\psi^{m}_{l}(\bold{x}) = \frac{1}{\left(2\pi\right)^{3/2}}e^{-|\bold{x}|^2 /2 \sigma^2}|\bold{x}|^l Y_l^m \left(\frac{\bold{x}}{|\bold{x}|}\right),
\end{equation}
which was first applied in a 3D molecular chemistry application \citep{10.5555/3295222.3295400,doi:10.1063/1.5023798}. In Eq. \eqref{solid:sup}, $Y_l^m$ denote the usual Laplacian spherical harmonics and $\sigma$ is the Gaussian width in units of the field grid size. The family of wavelets can then be generated by dilating the mother wavelet:
\begin{equation}\label{dil:sup}
\psi^{m}_{j, l}(\bold{x}) = 2^{-3 j}\psi^{m}_{l}(2^{-j} \bold{x}),
\end{equation}
spanning different dyadic scales, $2^{j}$, combined with varying values of the spherical harmonic of order $l$ to describe the angular information of the wavelet family, after we sum over the remaining index $m$. In this case, the WST coefficients are then given by:
\begin{align} \label{eq:WSTcoeff:sol}
 S_0 &= \langle |I(\bold{x})|^{q} \rangle, \nonumber \\
 S_1(j_1,l_1) &= \left\langle \left(\sum_{m=-l_1}^{m=l_1}|I(\bold{x}) \ast \psi^{m}_{j_1, l_1}(\bold{x}) |^{2}\right)^{\frac{q}{2}} \right\rangle, \\
 S_2(j_2,j_1,l_1) &= \left\langle \left(\sum_{m=-l_1}^{m=l_1}|U_1(j_1,l_1)(\bold{x}) \ast \psi^{m}_{j_2, l_1}(\bold{x}) |^{2}\right)^{\frac{q}{2}} \right\rangle \nonumber,
\end{align}
with
\begin{equation}
 U_1(j_1,l_1)(\bold{x}) = \left(\sum_{m=-l_1}^{m=l_1}|I(\bold{x}) \ast \psi^{m}_{j_1, l_1}(\bold{x}) |^{2}\right)^{\frac{1}{2}},
\end{equation}
which is obtained using a 3D solid harmonic mother wavelet \eqref{dil:sup} in Eq. \eqref{eq:WSTcoeff:power}. We briefly note that other wavelets considered in the literature are Morlet wavelets \citep{10.1093/mnras/staa3165,10.1093/mnras/stab2102}, bump-steerable wavelets \citep{PhysRevD.102.103506,2022arXiv220407646E} or the equivariant wavelet construction of Ref. \citep{Saydjari_2021}.

 The total number of essential WST coefficients can be further reduced, compared to Eq. \eqref{eq:WSTcoeff:power}, if we notice that the second order scales $j_2<j_1$, that is, scales smaller than the $1^{st}$ order convolution scale $j_1$, are practically filtered out and do not carry any extra information. This fact was indeed confirmed in the 2D weak lensing (WL) application by Ref. \citep{10.1093/mnras/staa3165} and was also subsequently adopted in our previous works \citep{PhysRevD.105.103534,PhysRevD.106.103509}. We will also work with only one second order angular scale, that is, for $l_2=l_1$ in Eq. \eqref{eq:WSTcoeff:sol}, following the choice originally adopted by the solid harmonic implementation of Ref. \citep{10.5555/3295222.3295400,doi:10.1063/1.5023798}. Even though orientations $l_2 \neq l_1$ are expected to be informative, this choice has been shown to be a good trade-off \citep{10.5555/3295222.3295400,doi:10.1063/1.5023798,PhysRevD.105.103534,PhysRevD.106.103509} and will be adopted in this work as well.
 
 The above choices determine the final number of produced WST coefficients, given as follows: for a certain total number of spatial scales $J$ and harmonic angular orientations $L$, we will have:
\begin{equation}
(j,l) \in([0,..,J-1,J],[0,..,L-1,L]),    
\end{equation}
giving rise to a total of 
\begin{equation}\label{Stotal}
S_0+S_1+S_2=1+(L+1)(J^2+3J+2)/2    
\end{equation}
WST coefficients up to $2^{nd}$ order. Since the dilations of the mother wavelet scale are chosen to be dyadic, $J\le\log_2({\rm NGRID})$, where NGRID is the resolution of the input field on each dimension. Finally, the width, $\sigma$, of the Gaussian in Eq. \eqref{solid:sup} and the power, $q$, in Eq. \eqref{eq:WSTcoeff:sol} are free parameters, whose values will be determined in the next section for our particular galaxy clustering application. 

To summarize, for a given choice of $J$, $L$, $q$ and $\sigma$, an input field $I(\bold{x})$ of resolution $\rm NGRID^{3}$ gives rise to the WST coefficients \eqref{Stotal} evaluated from Eq. \eqref{eq:WSTcoeff:sol}. We perform this evaluation using the publicly available package {\tt KYMATIO} \citep{2018arXiv181211214A}\footnote{Available in \url{https://www.kymat.io/}. We clarify that {\tt KYMATIO} evaluates the sum over all pixels of the input field, rather than the mean, which is the same up to a normalization, and thus exactly equivalent for parameter inference applications. We follow this version and, strictly speaking, we work with the sum over all pixels rather than the mean.}, as we will explain in the next section.

%%%%%%%%%%%%%%%%%%%%%%%%%%%%%%%%%%%%%%%%%%%%%%
\section{Dataset}\label{sec:Dataset}

\begin{figure}[ht]
\includegraphics[width=0.49\textwidth]{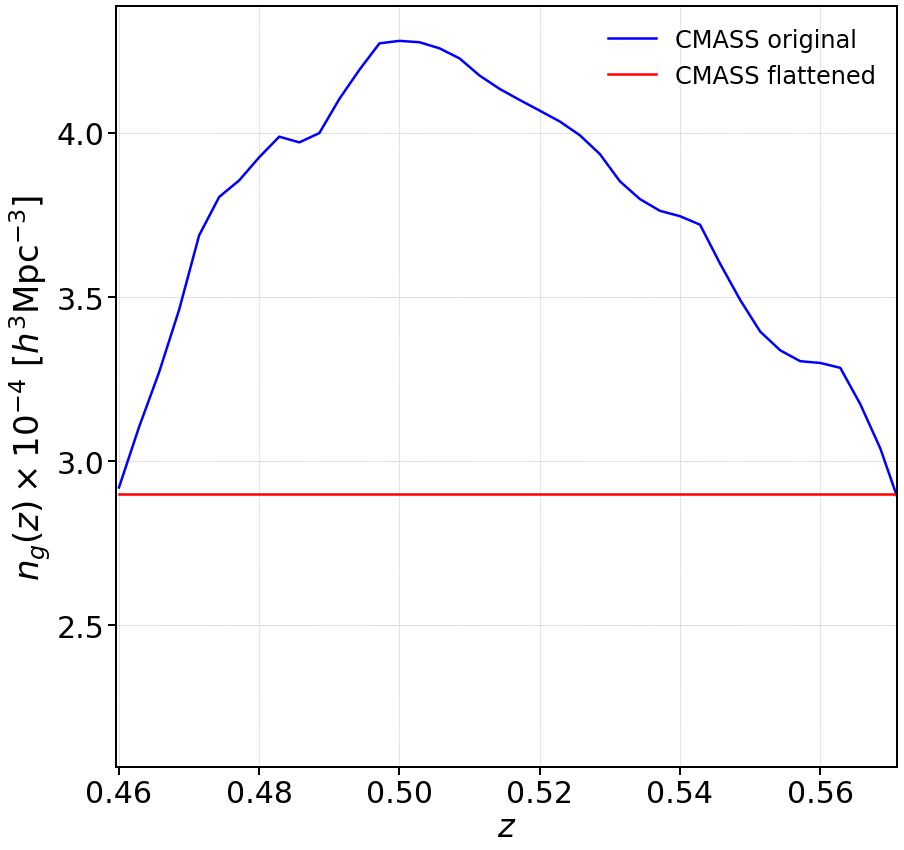}
\caption{The galaxy number density of the original CMASS sample as a function of redshift $z$ (blue), shown with the final downsampled version of constant number density, $\bar{n}_g = 2.9\times 10^{-4} \ h^3/{\rm Mpc}^3$ (red), that we work with in order to match the constant density profile of the \textsc{AbacusSummit} mocks.}
\label{fig:CMASSnz} 
\end{figure}

In this section we introduce the dataset that will be analyzed in this work.  This consists of Luminous Red Galaxies (LRGs) obtained from the twelfth data release (DR12) \citep{Alam_2015} of the Baryon Oscillation Spectroscopic Survey (BOSS), a part of Sloan Digital Sky Survey, SDSS-III \citep{Eisenstein_2011,Dawson_2012}, in particular the CMASS sample\footnote{All data are publicly available at \url{https://data.sdss.org/sas/dr12/boss/lss/}.}. Following our previous application \citep{PhysRevD.106.103509}, which was in turn aligned with the original analyses of BOSS data \citep{10.1093/mnras/stw1096,10.1093/mnras/stw3298}, we will work with each of the two subsamples obtained in the Northern (NGC) and the Southern Galactic Cap (SGC). If $X_{\rm NGC}$ and $X_{\rm SGC}$ denote the summary statistics evaluated from the Northern and Southern parts of the BOSS footprint, with angular area equal to $A_{\rm NGC}=6851 \ {\rm deg}^2$ and $A_{\rm SGC}=2525 \ {\rm deg}^2$, respectively, then we will always work with the weighted average 
\begin{equation}\label{eq:NplusS}
X_{\rm N+S} = \frac{\left(A_{\rm NGC}X_{\rm NGC}+A_{\rm SGC}X_{\rm SGC}\right)}{(A_{\rm NGC}+A_{\rm SGC})},
\end{equation}
where X in our analysis will be the data vector of the WST coefficients or the multipoles of the anisotropic correlation function of galaxies, as we will further explain in \S\ref{WST:model} and \S\ref{sec:Analysis}. Furthermore, we identify and work with the part of the sample with galaxy number density greater than $3\times 10^{-4} \ h^3/{\rm Mpc}^3$, corresponding to the redshift range $0.4613<z<0.5692$. In order to generate a sample with a constant density profile as a function of redshfit z, we further bin these galaxies into 50 linearly spaced z bins, and randomly downsample each bin such that the final outcome is a sample with a constant galaxy number density $\bar{n}_g = 2.9\times 10^{-4} \ h^3/{\rm Mpc}^3$. In Fig. \ref{fig:CMASSnz}, we show the original varying $n_g(z)$ of the sample, together with the final flattened profile that we are going to work with. The choice of the target 
 density $\bar{n}_g = 2.9\times 10^{-4} \ h^3/{\rm Mpc}^3$ is motivated by the mocks we use, and will be further explained in \S\ref{WST:model}. We also note that the choice to work with a flat density profile, which is meant to ensure a more accurate modeling of our density-dependent WST estimator, is different than our previous analysis \citep{PhysRevD.106.103509}, in which we worked with the original varying $n_g(z)$ in the range $0.46<z<0.60$. A similar choice has been made in other recent simulation-based reanalyses of BOSS data \citep{Zhai:2022yyk,2022MNRAS.509.1779L}.

%%%%%%%%%%%%%%%%%%%%%%%%%%%%%%%%%%%%%%%%%%%%%%
\section{Simulation-based forward model}\label{WST:model}

In this section we will describe the various ingredients used to construct our simulation-based model for the galaxy clustering and its summary statistics as a function of the cosmological and galaxy model parameters of interest. We begin with the suite of the \textsc{AbacusSummit} simulations used for the nonlinear modeling of the dark matter density and velocity fields, and then introduce the semi-analytical \ahod\ framework for populating the gravitationally bound dark matter halos with galaxies. Finally, we explain how we evaluate the WST coefficients and the 2-point correlation function multipoles of the galaxy mocks, in order to construct the training set for our emulator. 

%%%%%%%%%%%%%%%%%%%%%%%%%%%%%%%%%%%%%%%%%%%%%%
\subsection{The \textsc{AbacusSummit} simulations}
\label{sec:simulation}

\textsc{AbacusSummit} \citep[][]{2021Maksimova} is a suite of state-of-the-art cosmological N-body simulations that were run with the \textsc{Abacus} N-body code \citep{2019Garrison, 2021bGarrison}. Containing more than 150 high-accuracy and high-resolution simulations spanning almost 100 different cosmologies, it is capable of not only matching but also exceeding the simulation requirements of the Dark Energy Spectroscopic Instrument (DESI) survey \citep{2013arXiv1308.0847L, 2016DESI}. As a result, it is the ideal set to use in order to produce high-fidelity galaxy mocks for our BOSS CMASS simulation-based reanalysis.  We will exclusively work with the main (`base') set of cubic boxes with a side of length $2 \ \rm Gpc/h$, that evolved $6912^3$ dark matter particles with an individual mass equal to $2.1 \times 10^9 h^{-1}M_\odot$. 

In order to identify gravitationally bound dark matter halos in the simulations, the \textsc{AbacusSummit} uses a new efficient spherical-overdensity (SO) halo finder called {\sc CompaSO} \citep{2021Hadzhiyska}, which performs this task on-the-fly, and includes a series of improvements to avoid previously known challenges faced by halo finders, such as failure to identify structures close to larger halo centers or the blending of halos. Further details about \textsc{Abacus} and  {\sc CompaSO} can be found in the corresponding papers referenced above. 

%%%%%%%%%%%%%%%%%%%%%%%%%%%%%%%%%%%%%%%%%%%%%%
\subsubsection{The cosmology grid}\label{sec:grid}

Our cosmology grid consists of 85 simulations performed for different variations in the values of 8 cosmological parameters, which form the basis of our emulator and parameter inference setup. These parameters are: the baryon density $\omega_b = \Omega_b h^2$, the cold dark matter density $\omega_\mathrm{cdm} = \Omega_\mathrm{cdm}h^2$, the rms amplitude of linear density fluctuations at $8 \ \rm Mpc/h$ $\sigma_8$, the spectral tilt $n_s$, the running of the spectral tilt $\alpha_{run}$, the effective number of relativistic degrees of freedom $N_\mathrm{eff}$, and the dark energy equation of state parameters $w_0$ and $w_a$ ($w(a) = w_0+(1-a)w_a$), where $h=H_0/(100$ km s$^{-1}$Mpc$^{-1}$) is the dimensionless Hubble constant. Each one of the 85 simulations has been performed for the same fixed initial random phase, and with the value of the Hubble constant, $H_0$, chosen such that the comoving angular size of the sound horizon at last scattering, $\theta_{\star}$, is fixed to the corresponding value derived from measurements by the {\it Planck} satellite \citep{Planck2018}, $100 \theta_{\star}=1.041533$.

We refer to the different cosmologies using the naming scheme \textsc{cXXX}, where \textsc{XXX} ranges from 000 to 181. Details for each one of them are presented in the \textsc{AbacusSummit} website\footnote{\url{https://abacussummit.readthedocs.io/en/latest/cosmologies.html}}, with a visualization of the cosmological parameter grid shown in Fig.~1 of Ref. \citep{2022MNRAS.515..871Y} and its bounds listed in Table~\ref{tab:hodbounds}. 

We briefly describe the specifications of some selected cosmologies contained in the parameter grid. \textsc{c000} is a $\Lambda$CDM cosmology that corresponds to the parameters inferred by the {\it Planck} 2018 \citep{Planck2018} TT,TE,EE+lowE+lensing likelihood analysis, and which we pick as our fiducial cosmology from now on.

Furthermore, there are four secondary cosmologies exploring variations around the fiducial, \textsc{c001-004}, which we will use to validate the accuracy of our emulator in the next section. \textsc{c001} corresponds to the WMAP9 + ACT + SPT cosmology \citep[][]{2017Calabrese}, while \textsc{c002} is a wCDM cosmology with $w_0 = -0.7$ and $w_a = -0.5$. Finally, \textsc{c003} is a cosmology with higher $N_\mathrm{eff} = 3.7$, and \textsc{c004} has a low clustering amplitude given by $\sigma_8$ = 0.75. 

The \textsc{AbacusSummit} also contains additional simulations that vary each one of the 8 cosmological parameters, in turn, and in a step-wise fashion around the fiducial \textsc{c000}, while keeping the rest fixed, in order to enable the evaluation of first-order derivatives for summary statistics. This linear derivative grid consists of cosmologies \textsc{c100-126}, which were used to construct the Taylor expansion approximation we adopted in Ref. \citep{PhysRevD.106.103509}. 
Cosmologies \textsc{c130-181} form a broad parameter grid that provides a wider coverage of the 8-dimensional target parameter space and enables the training of emulators. Further details on the motivations behind the choice of these cosmologies and the parameter ranges can be found in Ref. \citep{2021Maksimova} and the \textsc{AbacusSummit} website.

Lastly, in order to quantify the effects of sample variance and potential errors introduced when training at a single phase, a second set of simulations with the same specifications has been run for 24 additional random realizations of the \textsc{c000} fiducial cosmology. The phase information is labeled as ph000-024. In the next sections we will describe how we used both of the above sets in order to accurately train our emulator for the vector of WST coefficients and the multipoles of the 2-point correlation function. 

%%%%%%%%%%%%%%%%%%%%%%%%%%%%%%%%%%%%%%%%%%%%%%
\subsection{The Halo Occupation Distribution (HOD)}\label{subsec:model}

The galaxy--halo connection model we use to generate the galaxy mocks for our forward model is known as the Halo Occupation Distribution (HOD) (see, e.g., Refs. \citep{2005Zheng, Zheng_2007}), which is a probabilistic model that populates dark matter halos with galaxies through a set of empirical formulas conditioned on halo properties. For a Luminous Red Galaxy (LRG) sample such as CMASS, the HOD is well approximated by a vanilla model given by:
\begin{align}
    \bar{n}_{\mathrm{cent}}^{\mathrm{LRG}}(M) & = \frac{1}{2}\mathrm{erfc} \left[\frac{\log_{10}(M_{\mathrm{cut}}/M)}{\sqrt{2}\sigma}\right], \label{equ:zheng_hod_cent}\\
    \bar{n}_{\mathrm{sat}}^{\mathrm{LRG}}(M) & = \left[\frac{M-\kappa M_{\mathrm{cut}}}{M_1}\right]^{\alpha}\bar{n}_{\mathrm{cent}}^{\mathrm{LRG}}(M),
    \label{equ:zheng_hod_sat}
\end{align}
where the five parameters characterizing the model are $M_{\mathrm{cut}}, M_1, \sigma, \alpha, \kappa$. The parameter $M_{\mathrm{cut}}$ defines the minimum halo mass to host a central galaxy, $M_1$ sets the typical halo mass that hosts one satellite galaxy, $\sigma$ characterizes the steepness of the error function upturn in the number of central galaxies, $\alpha$ is the power-law index on the number of satellite galaxies, and $\kappa M_\mathrm{cut}$ controls the minimum mass of a halo that can host a satellite galaxy. We have also added a modulation term $\bar{n}_{\mathrm{cent}}^{\mathrm{LRG}}(M)$ to the satellite occupation function to disfavor satellites from halos without centrals. This term represents a model choice and is inconsequential for the conclusions of this work. 

The HOD model does not only provide predictions for the number of galaxies populating each halo, but it also determines the positions and velocities of these galaxies. In the case of the central galaxies, their positions and velocities match the ones of the halo center-of-mass (the L2 subhalo when working with {\sc CompaSO}), while the satellites are randomly assigned to halo particles with uniform weights, each satellite inheriting the position and velocity of its host particle. Note that we do not impose any satellite radial profile in this model. 

We also include a motivated HOD extension known as velocity bias, which biases the velocities of the central and satellite galaxies relative to their host halos and particles. This is shown to to be a necessary ingredient in modeling BOSS LRG redshift-space clustering on small scales \citep[e.g.][]{2015aGuo, 2021bYuan}. The velocity bias has also been identified in hydrodynamical simulations and measured to be consistent with observational constraints \citep[e.g.][]{2022Yuan, 2017Ye}. 
\newline We parametrize the velocity bias through two additional HOD parameters: 
\begin{itemize}
    \item \texttt{$\alpha_\mathrm{vel, c}$} controls the peculiar velocity of a central galaxy relative to the halo center, and is called the central velocity bias parameter. For instance, a value of $\alpha_\mathrm{vel, c} = 0$  indicates that centrals perfectly track the velocity of halo centers.
    \item \texttt{$\alpha_\mathrm{vel, s}$}, the satellite velocity bias, is the equivalent parameter for the satellite galaxies, modulating how their peculiar velocities deviate from those of the local dark matter particles.
    A value of $\alpha_\mathrm{vel, s} = 1$ indicates that satellites perfectly track the velocity of the underlying dark matter particles. 
\end{itemize}

Furthermore, we do not include the effects of assembly bias in our analysis, given that they typically manifest in smaller scales than the ones we consider, as we clarify below. We additionally check our cosmology recovery against a galaxy mock that contains galaxy assembly bias in section~\ref{sec:Abactests}. Nevertheless, we acknowledge the lack of robust galaxy assembly bias modeling as a potential systematic.
We reserve an analysis extending to smaller scales for a follow-up investigation. For a detailed discussion on the effects on assembly bias on cosmological analyses of BOSS CMASS, readers are referred to Ref. \citep{10.1093/mnras/stab235}. 

For computational efficiency, we adopt the highly optimized \ahod\ implementation, which significantly speeds up the HOD calculation per HOD parameter combination \citep[][]{2021bYuan}. The code is publicly available as a part of the \textsc{abacusutils} package at \url{https://github.com/abacusorg/abacusutils}. Example usage can be found at \url{https://abacusutils.readthedocs.io/en/latest/hod.html}. In order to match the clustering of CMASS in the redshift range $0.4613<z<0.5692$, we produce cubic galaxy mocks (of side $2 \ \rm Gpc/h$) at redshift $z=0.5$.

To summarize, the HOD model used in this analysis is fully parameterized by 7 parameters, $M_{\mathrm{cut}}, M_1, \alpha$, $\alpha_\mathrm{vel, c}$, $\alpha_\mathrm{vel, s}$, $\kappa$ and $\sigma$. 

%%%%%%%%%%%%%%%%%%%%%%%%%%%%%%%%%%%%%%%%%%%%%%
\subsection{Survey Geometry}\label{subsec:cutsky}

The \textsc{AbacusSummit}  galaxy mocks that we produce with \ahod\ come in a periodic cubic geometry with a side equal to $2 \ h^{-1}$Gpc, at output redshift $z=0.5$, as we previously discussed. This configuration is different from the non-trivial survey geometry of the CMASS sample that we will analyze in this work, which was introduced in \S\ref{sec:Dataset}. When working with conventional statistics, the effect of a non-trivial survey geometry can be usually captured with a model. In the case of the galaxy power spectrum, for example, the prediction for a periodic configuration is convolved with the Fourier transform of the survey mask \citep{1994ApJ...426...23F,10.1093/mnras/stw1096,10.1093/mnras/stu1051,10.1093/mnras/stw2576,10.1093/mnras/stw3298} or, equivalently, the prediction from the masked data can be de-convolved \citep{Beutler_2021}. Given that no such model is available for the WST, which is sensitive to the survey geometry through the successive wavelet convolutions, we proceed to directly cut the Abacus cubes into the exact 3D shape of the BOSS CMASS data, as we did in our previous work \citep{PhysRevD.106.103509}\footnote{Alternatively, one could consider using modern inpainting techniques \citep{Mohammad:2021ylk}.}. Specifically, each cubic mock of redshift $z=0.5$ is downsampled to a constant number density $\bar{n}_g = 2.9\times 10^{-4} \ h^3/{\rm Mpc}^3$ and is then fed as input into the public code \textsc{make\_survey} \citep{White:2013psd}\footnote{Available at \url{https://github.com/mockFactory/make_survey}.}. Using the real-space Cartesian positions and velocities for each galaxy at $z=0.5$, the CMASS angular footprint, as well as the parameters for each cosmology \textsc{cXXX}, \textsc{make\_survey} transforms the original cubic mocks into galaxy catalogs with sky coordinates right ascension (RA), declination (DEC), and redshift z that exactly match the 3D geometry of the observed CMASS sample in the target range $0.4613<z<0.5692$, with the redshift-space distortion (RSD) implemented along the radial direction. The procedure is repeated twice for each mock in order to produce separate samples for NGC and SGC, respectively. As in Ref. \citep{PhysRevD.106.103509}, we confirm the robustness of this procedure by evaluating the power spectra of both the original cubic and the final cut-sky mocks and by making sure they remain unchanged, up to sample variance error. 

%%%%%%%%%%%%%%%%%%%%%%%%%%%%%%%%%%%%%%%%%%%%%%
\subsection{Summary statistic evaluation}\label{subsec:summary_statistic}

Having laid out the procedure to generate realistic galaxy mocks that resemble the footprint of the CMASS sample as a function of the cosmological and HOD parameters, we now proceed to explain how we evaluate the summary statistics of interest, starting with the WST coefficients. 

%%%%%%%%%%%%%%%%%%%%%%%%%%%%%%%%%%%%%%%%%%%%%%
\subsubsection{WST}\label{subsec:wsteval}

The quantity of interest for the density-dependent WST estimator is the fractional overdensity field of galaxies, which we evaluate with the following procedure: the sky coordinates RA, DEC and z of each galaxy in each sample (be it either the cut-sky mocks or the CMASS data) are converted to comoving Cartesian coordinates (x,y,z), always assuming a fiducial flat $\Lambda$CDM cosmology with $\Omega_m=0.3152,h=0.6736$ (corresponding to the \textsc{Abacus} cosmology c000). Each sample is then embedded into the smallest possible 3D cube for this task, which we determine with the public package {\tt nbodykit}\footnote{\url{https://nbodykit.readthedocs.io/en/latest/index.html}.}, and which is found to have a comoving side $L=2700$ $\rm Mpc/h$ for the range $0.4613<z<0.5692$. When working with spectroscopic data in sky coordinates, the relevant quantity is the (weighted) fractional overdensity of galaxies, also known as the Feldman-Kaiser-Peacock (FKP) field, $F(r)$ \citep{1994ApJ...426...23F}, given by:
\begin{equation}\label{fkpfield}
F(\bold{r})=\frac{w_{\rm FKP}(\bold{r})}{I_2^{1/2}}\left[w_c(\bold{r})n_g(\bold{r})-\alpha_r n_s(\bold{r}) \right],
\end{equation}
which can be evaluated on a 3D Cartesian grid. The quantities $n_g(\bold{r})$ and $n_s(\bold{r})$ in Eq. \eqref{fkpfield} denote the observed number density of the galaxies compared to the one of a random, unclustered, catalogue, respectively, with the latter containing $\alpha_r$ times more objects. Furthermore, the BOSS dataset is accompanied by a set of systematic weights given by Refs. \citep{10.1093/mnras/stw1096,10.1093/mnras/stw3298}:
\begin{equation}\label{eq:wc}
w_c(\bold{r}) = \left(w_{\rm rf}(\bold{r})+ w_{\rm fc}(\bold{r}) - 1.0 \right)w_{\rm sys}(\bold{r}), 
\end{equation}
in which a fiber collision weight, $w_{\rm fc}$, a systematics weight, $w_{\rm sys}$ and  a redshift failure weight, $w_{\rm rf}$, are combined to account for the various incompletenesses of the observed sample. Aiming to ensure optimal recovery of small-scale information from the galaxy power spectrum, we traditionally also define the FKP weight \citep{1994ApJ...426...23F}:
\begin{equation}\label{eq:fkpweight}
w_{\rm FKP}(\bold{r})=\left[1+\bar{n}_g(\bold{r})P_0\right]^{-1}, 
\end{equation}
where $P_0=10^{-4} \ {\rm Mpc}^3/h^3$, and where the normalization factor 
\begin{equation}\label{eq:Inorm}
I_2 = \int d^3\bold{r} \ w_{\rm FKP}^2 (\bold{r}) \langle w_c(\bold{r})n_g(\bold{r}) \rangle ^2
\end{equation}
is defined in Eq. \eqref{fkpfield}, with respect to the amplitude of the regular galaxy power spectrum of a uniform sample. In addition to the values of the systematic weights \eqref{eq:wc} for each observed galaxy, the public BOSS release also includes random catalogues matching the same selection function and footprint of the survey, in order to enable the evaluation of $n_s(\bold{r})$ in Eq. \eqref{fkpfield}. Out of the various options available, we choose to work with the random catalogue corresponding to $\alpha_r=50$, which is the commonly adopted choice in the literature \citep{10.1093/mnras/stw1096,10.1093/mnras/stw3298}. When working with a sample that does not possess incompleteness weights, as is the case for the galaxy mocks, Eq. \eqref{fkpfield} merely corresponds to the regular galaxy overdensity field, but in a non-trivial survey geometry. In order to evaluate the random density field $n_s(\bold{r})$ in this case, we similarly generate a random cubic sample with 50 times higher number density than the original mocks, and then subject it to the same cut-sky procedure that we described in the previous subsection.

We should note, at this point, that in order to convert the sky coordinates of the galaxies in the CMASS sample into comoving Cartesian ones, we assumed a (potentially incorrect) flat $\Lambda$CDM cosmology corresponding to $\Omega_m=0.3152,h=0.6736$. This assumption introduces an error usually referred to as the Alcock–Paczynski (AP) distortion \citep{1979Natur.281..358A}. To account for this effect in our model, we always assume the above same cosmology when converting the coordinates RA, DEC and z of the \textsc{Abacus} mocks back into comoving ones, even though the true cosmological parameters for each one of the \textsc{cXXX} boxes is actually known (and were used to convert the original cubes into cut sky mocks). The procedure is the equivalent one to the power spectrum re-scalings usually applied in order to account for the AP effect in traditional BOSS analyses \citep{Ivanov_2020,d_Amico_2020,Philcox_2020,PhysRevD.105.043517,Chen_2022,Zhang_2022,2022MNRAS.509.1779L}, that we also adopted in Ref. \citep{PhysRevD.106.103509}.

Finally, Eq. \eqref{fkpfield} with the corresponding systematic weights from Eq. \eqref{eq:wc} (or unweighted) and the FKP weight \eqref{eq:fkpweight}, can be combined with the random catalogues in order to enable the evaluation of the final FKP density field from the CMASS data (\textsc{Abacus} mocks). Following the choices adopted in our previous WST BOSS analysis \citep{PhysRevD.106.103509}, we resolve the field on a mesh of resolution ${\rm NGRID=270}$, using the Triangular Shaped Cloud (TSC) mass assignment scheme \citep{hockney1981computer}, and work with a Gaussian width $\sigma=0.8$ in Eq. \eqref{eq:WSTcoeff:sol}, such that the smallest density cell corresponds to a scale of length $8 \ h^{-1}$Mpc on the side. This FKP field serves as input into the WST network \eqref{eq:WSTcoeff:sol} in order to evaluate the relevant WST coefficients. Combining the above choices with $J=4$ scales, $L=4$ orientations, and $q=0.8$ (as in Ref. \citep{PhysRevD.106.103509}), we obtain the target data vector of 76 WST coefficients from Eq. \eqref{eq:WSTcoeff:sol}. The evaluation is performed with {\tt KYMATIO} \citep{2018arXiv181211214A}, using our modified version for an application to a masked galaxy field (as explained further in Appendix A of Ref. \citep{PhysRevD.106.103509}). We note that the overall evaluation of the WST coefficients out of an original \textsc{Abacus} cubic mock through the pipeline described above takes about 60 seconds per core when the WST evaluation is GPU-accelerated.

%%%%%%%%%%%%%%%%%%%%%%%%%%%%%%%%%%%%%%%%%%%%%%
\subsubsection{2-point correlation function}

In order to have a benchmark that will allow us to assess the performance of the WST compared to standard cosmological analyses, we also evaluate the 2-point correlation function of galaxies. In particular, if by $\xi(s,\mu_s)$ we denote the 2D anisotropic correlation function of galaxies as a function of redshift space separation $s$, then its multipoles, $\xi_{\ell}(s)$, can be extracted through the usual expansion
\begin{equation}\label{eq:xi2D}
\xi(s,\mu_s) = \sum_l \xi_{\ell}(s) L_{\ell}(\mu_s)
\end{equation}
in a basis of Legendre Polynomials $L_{\ell}(\mu_s)$, which then gives
\begin{equation}\label{eq:xiell}
 \xi_{\ell}(s) = \left(2\ell+1\right)\int_0^{1} \xi(s,\mu_s)L_{\ell}(\mu_s) d\mu_s.
\end{equation}

For a sample of galaxies in sky coordinates, which we are working with in this analysis, $\mu_s = \hat{s}\cdot \hat{r}$, where the radial anisotropy direction $\hat{r}$ is the line-of-sight (as opposed to one of the Cartesian axes direction when working with a periodic box). We choose to work with the two lowest non-vanishing multipoles, $\ell=\{0,2\}$, which we evaluate with the public code \textsc{Pycorr}\footnote{\url{https://github.com/cosmodesi/pycorr}.}, which is a wrapper for \textsc{Corrfunc} \citep{2020MNRAS.491.3022S}\footnote{\url{https://corrfunc.readthedocs.io/en/master/index.html}}, using the  Landy-Szalay (LS) estimator \citep{Landy:1993yu} with 241 linearly spaced angular bins in $-1<\rm \mu<1$. For the spatial separation, we adopt a differential binning strategy, which is the following: for the monopole, we use 10 linearly spaced bins centered between $10<s<56\ \rm Mpc/h$ followed by 6 bins for scales $67<s<142\ \rm Mpc/h$, while for the quadrupole we downsample the above binning scheme by a factor of 2. This choice, which corresponds to a total of 24 bins, was found to deliver the optimal trade-off between large-scale noise due to cosmic variance and the ability to capture the full shape of the correlation function. We also note that this binning scheme is still finer that the one chosen for the WST, in order to ensure a fair comparison between the performance of the two statistics.

This evaluation can be straightforwardly performed using the sky positions of the CMASS galaxy sample (or the simulated mocks), as well as the ones of the accompanied random catalogues, as input to \textsc{Corrfunc}. For the conversion of the sky coordinates into comoving ones, we adopted the same fiducial cosmology as discussed in \S\ref{subsec:wsteval} for the WST, in connection to the AP effect. The choice of $s_{min}$ matches the minimum scale accessed by the WST, for which we used a cell of grid size $8 \ \rm Mpc/h$ as explained above, in order to ensure a fair comparison. (Further discussion on the minimum scale cut can be found in Appendix \S\ref{Appsec:kmax} ).

%%%%%%%%%%%%%%%%%%%%%%%%%%%%%%%%%%%%%%%%%%%%%%
\subsection{Emulator}\label{subsec:Emu}

\begin{table}
    \centering
    \begin{tabular}{|l|c|}
        \hline
        \hline
        Parameter                  &Bounds\\
        \hline
        $\omega_b$              &[0.0207, 0.0243]\\
        $\omega_c$              &[0.1032, 0.14]\\
        $\sigma_8$              &[0.687, 0.938]\\
        $n_s$                   &[0.901, 1.025]\\
        $a_{\rm run}$               &[-0.038, 0.038]\\
        $N_{\rm eff}$               &[2.1902, 3.9022]\\
        $w_0$                   &[-1.27, -0.70]\\
        $w_a$                   &[-0.628, 0.621]\\
        \hline
        $\log_{10} M_{\rm cut}$          &[12.4, 13.3]\\
        $\log_{10} M_1$              &[13.0, 15.0]\\
        $\sigma$                &[0.001, 1.0]\\
        $\alpha$                &[0.5, 1.5]\\
        $\kappa$                &[0.0, 8]\\
        $\alpha_\mathrm{c}$     &[0.0, 0.8]\\
        $\alpha_\mathrm{s}$     &[0.0, 1.5]\\
        \hline        
    \end{tabular}
    \caption{Priors bounds used to generate the cosmology + HOD training set of our emulator. Units of mass are in $h^{-1}M_\odot$. The HOD values are roughly centered on results from Ref. \citep{2022MNRAS.515..871Y}.
    }
    \label{tab:hodbounds}
\end{table}
After explaining the steps to go from the original \textsc{AbacusSummit} simulations to realistic galaxy mocks resembling the properties of the CMASS sample, we now lay out the details of our emulation scheme for the cosmological dependence of the target summary statistics.

Emulators refer to parametrized surrogate models for the cosmological dependence of a summary statistic used to interpolate sparse likelihood evaluations. The emulator replaces the expensive likelihood calls with the much cheaper emulator model calls, thus enabling a much faster sampling at the cost of introducing additional errors in the model training. Such schemes have become increasingly popular in simulation-based cosmological analyses with the advent of fast yet flexible machine learning models such as neural nets and Gaussian processes, with a series of successful cosmology applications in recent years (see e.g. Refs. \citep{2009Heitmann, 2010Lawrence, 2014Heitmann, 2021PhRvD.103l3525R, 2019Zhai, 2022Zhai, 2021Lange, 2021Kobayashi, 2022MNRAS.515..871Y, 2023Yuan}).

\begin{figure}
    \includegraphics[width=0.47\textwidth]{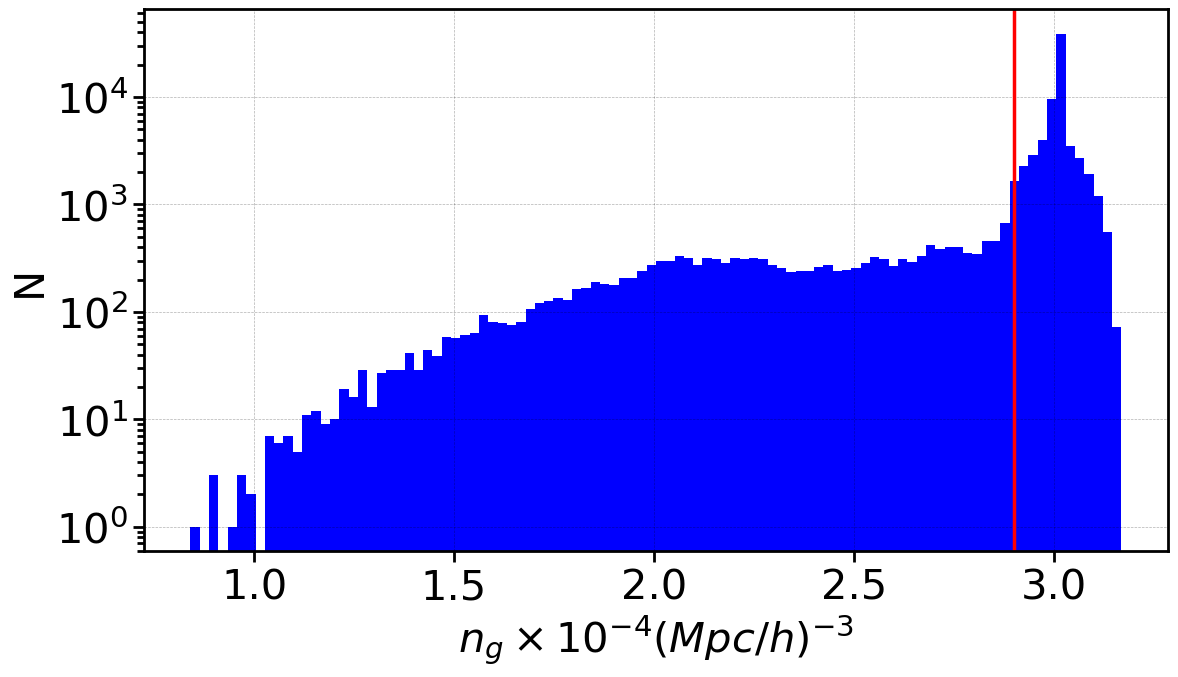}
    \vspace{-0.4cm}
    \caption{A histogram of the distribution of galaxy number densities of the galaxy mocks forming our original emulator parameter grid. Mocks with number densities lower than the cut-off $\bar{n}_g = 2.9\times 10^{-4} \ h^3/{\rm Mpc}^3$ (red vertical line) are discarded, while the rest are downsampled to exactly match this value, in order to ensure a robust modeling of the density-dependent WST estimator since this is the constant density value used in the \textsc{AbacusSummit} mocks. The outcome of this procedure forms the final emulator training set consisting of 151474 mocks, i.e. those lying on the right of the  red vertical line in the histogram.}
    \label{fig:Nmocks}
\end{figure}

To generate the training and test set, we forward model the final summary statistics (WST coefficients and 2-point correlation function) across 85 cosmologies and 2700 HOD variations at each cosmology, creating an initial set of 229500 mocks. The cosmology grid is described in \S\ref{sec:grid} and spans the $w$CDM+$N_\mathrm{eff}$+running space around {\it Planck} 2018 values \citep{Planck2018}. We leave out the four secondary cosmologies \textsc{c001-004} as out-sample tests. The HODs are sampled in a Latin Hypercube with flat bounded priors along each HOD parameter direction. The bounds for all parameters are summarized in Table~\ref{tab:hodbounds}. For each cosmology and HOD, we generate the periodic galaxy mocks according to the steps described in \S\ref{sec:simulation} and \S\ref{subsec:model}, discard the mocks that have number density lower than $2.9\times 10^{-4} \ h^3/{\rm Mpc}^3$, and randomly downsample the galaxies of the other mocks in order to exactly match the target density $\bar{n}_g = 2.9\times 10^{-4} \ h^3/{\rm Mpc}^3$. The value of this density cut-off allows us to retain a significant portion of the original collection of 229500 mocks, while discarding HOD configurations resulting in very low number densities, as shown in Fig.~\ref{fig:Nmocks}. We end up retaining 151474 cubic mocks with number density $\bar{n}_g = 2.9\times 10^{-4} \ h^3/{\rm Mpc}^3$, each one of which is cut to give two independent cut sky galaxy samples for NGC and SGC, respectively, as explained in \S\ref{subsec:cutsky}. We extract the summary statistics (WST and 2-point function) out of each one of them, as explained in \S\ref{subsec:summary_statistic}, and finally obtain the corresponding sky-averaged quantities according to Eq.~(\ref{eq:NplusS}), which form our final emulator training+test set. 

For the emulator, we adopt a fully connected neural network as our surrogate model. For the emulation of WST, we adopt a network of 3 layers as our fiducial model, with 300 nodes in each layer and a Gaussian Error Linear Unit (GELU) activation function. We train the network with the Adam optimizer and a mean squared loss function taking the diagonal terms of the CMASS WST covariance matrix (the evaluation of which is explained in \S\ref{subsec:likelihood}) as weights. We follow a mini-batch procedure and conduct cross-validation throughout the training process. 

\begin{figure}
    \includegraphics[width=0.5\textwidth]{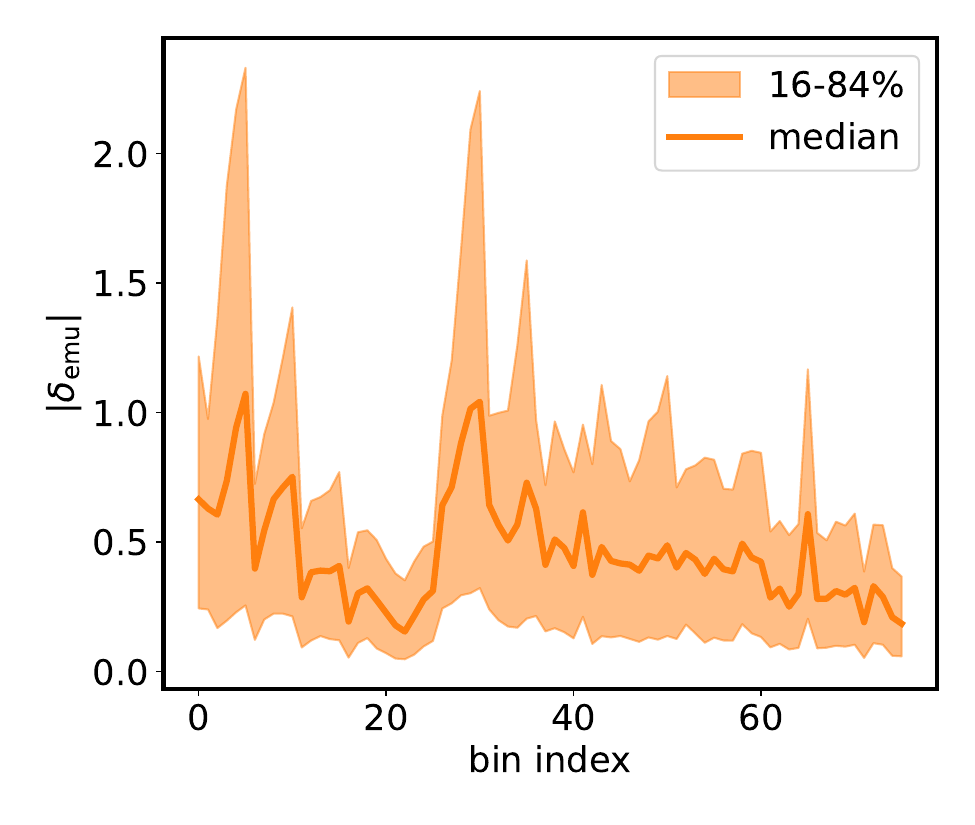}
    \vspace{-0.4cm}
    \caption{The median WST emulator error tested on the four leave-out cosmologies as a function of bin index of the WST coefficients vector. The $y$-axis denotes the emulator error normalized by the CMASS $1\sigma$ uncertainty. }
    \label{fig:emu_err_med}
\end{figure}

\begin{figure*}
    \includegraphics[width=0.95\textwidth]{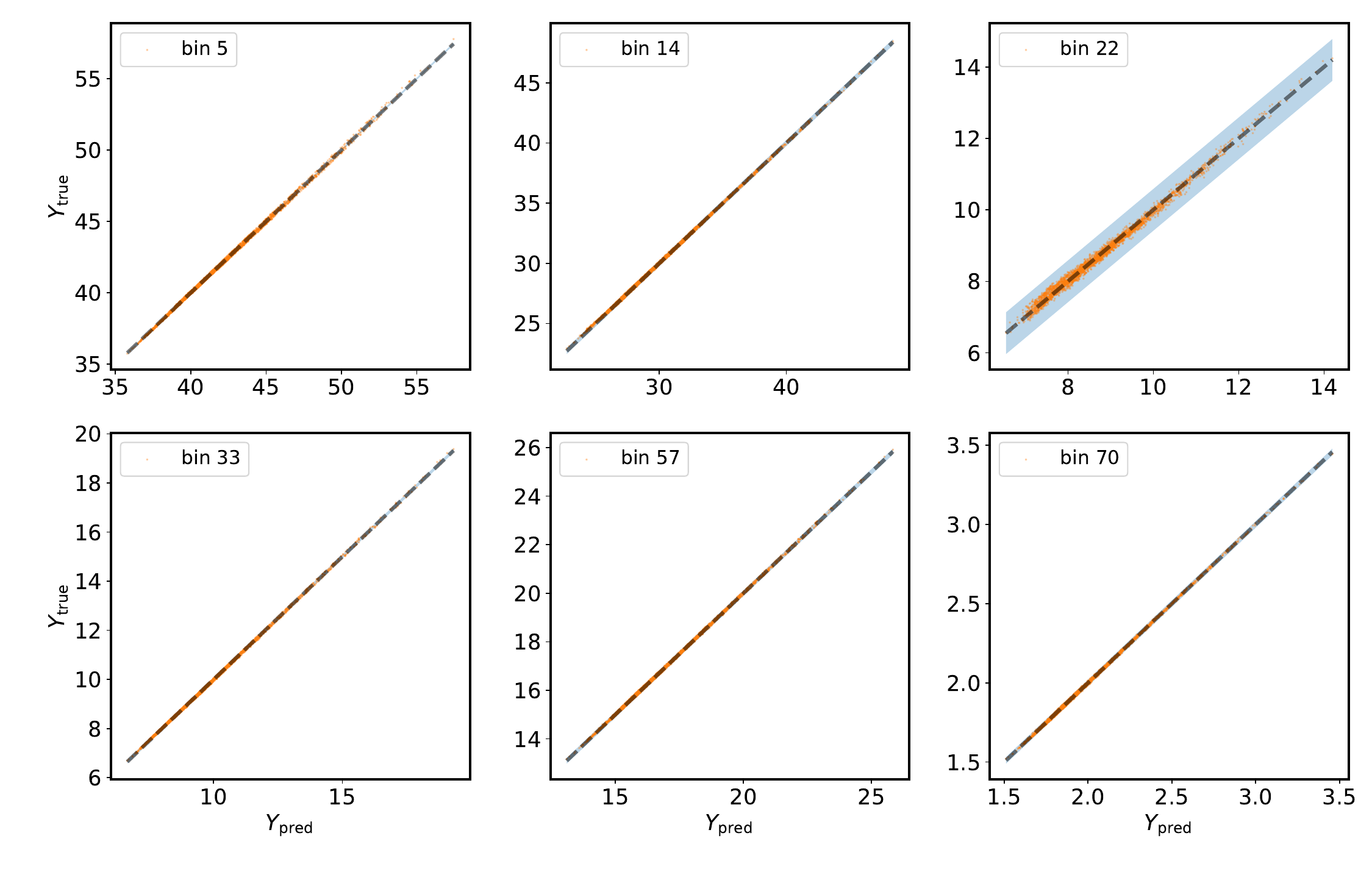}
    \vspace{-0.4cm}
    \caption{The bias of the WST emulator tested on the four leave-out cosmologies for six randomly selected coefficients (bins) of the WST data vector. The legend shows the WST bin indices. The orange scatter points showcase the true and predicted values of the WST coefficients for each one of the 4000 leave-out tests, whereas the blue band corresponds to the $1\sigma$ uncertainty of the CMASS WST measurement.}
    \label{fig:emu_err_bias}
\end{figure*}

We visualize the final WST emulator performance in Figs. \ref{fig:emu_err_med} and \ref{fig:emu_err_bias}. Specifically, Fig.~\ref{fig:emu_err_med} summarizes the emulator error relative to the CMASS uncertainty, $\delta_\mathrm{emu}$, as a function of WST bin indices. The errors are computed on 1000 HODs (sampled from the prior) at the four out-sample test cosmologies. The bins with the largest relative errors are the ones probing the largest spatial scales, which are more susceptible to cosmic variance and thus exhibit a larger dispersion in their values. We report a mean $|\delta_\mathrm{emu}|$ of 0.51, suggesting that the emulator error is overall sub-dominant relative to the measurement uncertainties. Fig.~\ref{fig:emu_err_bias} compares the true WST values and the emulator predicted values across all test cases for a few selected WST coefficients of the data vector (i.e. bins). The orange points show the emulator predictions for the respective coefficients for each one of the 4000 leave-out test cases, whereas the blue band shows the measurement uncertainties. The dashed line shows the $Y_\mathrm{pred} = Y_\mathrm{true}$ line. We see no sign of bias in the emulator prediction. Lastly, we repeat the above steps for the WST in order to create the corresponding emulator for the multipoles of the 2-point correlation function using the same training set.

%%%%%%%%%%%%%%%%%%%%%%%%%%%%%%%%%%%%%%%%%%%%%%
\section{Analysis}\label{sec:Analysis}

In this section, we lay out the details of how we will use our forward model for the galaxy clustering in order to perform a likelihood analysis of the BOSS data. We start with a description of the adopted likelihood we sample from and then explain our steps to validate the robustness of our pipeline.

%%%%%%%%%%%%%%%%%%%%%%%%%%%%%%%%%%%%%%%%%%%%%%
\subsection{Likelihood modeling}\label{subsec:likelihood}

Having laid out our methodology on how to forward model the cosmological dependence of the WST coefficients, as well as on how to extract the corresponding prediction from the data, we now explain our strategy for combining these necessary ingredients to perform a likelihood analysis of the BOSS dataset. Consider $\bold{X}$ to be the summary statistic of interest, which in our analysis denotes either the vector of WST coefficients or the multipoles of the correlation function (or their combination). Assuming $\bold{X}$ is Gaussian-distributed, as we confirm in Appendix \S\ref{Appsec:Gaussianity}, its likelihood, $\mathcal{L}(\theta|\bold{d})$, will then follow the familiar form:
\begin{equation}\label{eq:LogL}
\log \mathcal{L}(\theta|\bold{d}) = -\frac{1}{2} \left[\bold{X}_\bold{d}-\bold{X}_t(\theta)\right]^{\rm T} C^{-1}\left[\bold{X}_{\bold{d}}-\bold{X}_t(\theta)\right] + {\rm const.},
\end{equation} 
with $\bold{X}_\bold{d}$ being the value of the estimator evaluated from the BOSS CMASS dataset ${\bf d}$, that we will analyze in order to infer the set of parameters $\theta$. Furthermore, $C$ in Eq. \eqref{eq:LogL} denotes the covariance matrix of $\bold{X}$, which can be decomposed as in Ref. \citep{2022MNRAS.515..871Y}:
\begin{equation}\label{eq:covmatall}
C = C_{\bold{d}}+C_{\rm emu}+C_{\rm phase}.
\end{equation} 
The first term in Eq. \eqref{eq:covmatall}, $C_{\bold{d}}$, represents the usual contribution from the sample variance of the CMASS dataset ${\bf d}$, given by:
\begin{equation}\label{eq:covmat}
C_{\bold{d}} = \frac{1}{\rm N_{mocks} - 1}\sum_{\rm k=1}^{\rm N_{\rm mocks}} \left(\bold{X}^k_P-\bar{\bold{X}}_P\right)\left(\bold{X}^k_P-\bar{\bold{X}}_P\right)^{\rm T},
\end{equation} 
which we evaluate using $N_{\rm mocks}=2048$ realizations of the \textsc{Patchy} mocks (to be described in \S\ref{subsec:Patchy}), and with $\bar{\bold{X}}_P$ being the mean prediction over the $N_{\rm mocks}$. Furthermore, we follow Ref. \citep{2022MNRAS.515..871Y} and consider two extra contributions to the overall error budget, which reflect additional sources of uncertainty arising from our forward model and are essential for a reliable interpretation of our analysis. In particular, $C_{\rm emu}$ quantifies the residual emulator error evaluated (at fixed phase ph000) by averaging over the $4\times 1000=4000$ hold-out test errors, generated from \textsc{c001}-\textsc{c004} (as intoduced in \S\ref{sec:simulation}). These hold-out tests and their results will be described in detail in \S\ref{sec:Abactests}.

Furthermore, $C_{\rm phase}$ is meant to capture the effect of training using mocks at a fixed phase, rather than the average over many random realizations. To mitigate this effect, we make use of 24 additional simulations initialized at phases ph001-ph024, for the fiducial \textsc{c000} cosmology and a fixed HOD (corresponding to the best-fit values from~\citep{2022MNRAS.515..871Y}). We then apply the following phase correction to the data vector:
\begin{equation}\label{eq:phasecorr}
\bold{X}_{\rm smooth} = \bold{X}_{\rm ph000}\left[\frac{\bar{\bold{X}}}{\bold{X}_{\rm ph000}}\right],
\end{equation} 
where $\bold{X}_{\rm ph000}$ is the original emulator prediction, trained at fixed phase, and the term inside the brackets denotes the fractional correction evaluated over the 25 random realizations for the cosmology \textsc{c000}. Even though Eq. \eqref{eq:phasecorr} assumes that this phase effect is cosmology-independent, it was found in Ref.~\citep{2022MNRAS.515..871Y} to be sufficient for the mitigation of cosmic variance to the emulator predictions, and we adopt it here as well. Equivalently, one could explicitly evaluate an error term, $C_{\rm phase}$, using the 25 random phase realizations, as in Eq. \eqref{eq:covmat}. We have tried both approaches and have found minimal differences between the results of the corresponding likelihood analyses.
It is straightforward to see that in the limit of perfect emulator accuracy, these two additional terms would vanish, and Eq. \eqref{eq:covmatall} would reduce to its usual expression capturing only the cosmic variance \eqref{eq:covmat}, but we will find that these effects are not negligible. 

\begin{figure*}[ht]
\includegraphics[width=0.75\textwidth]{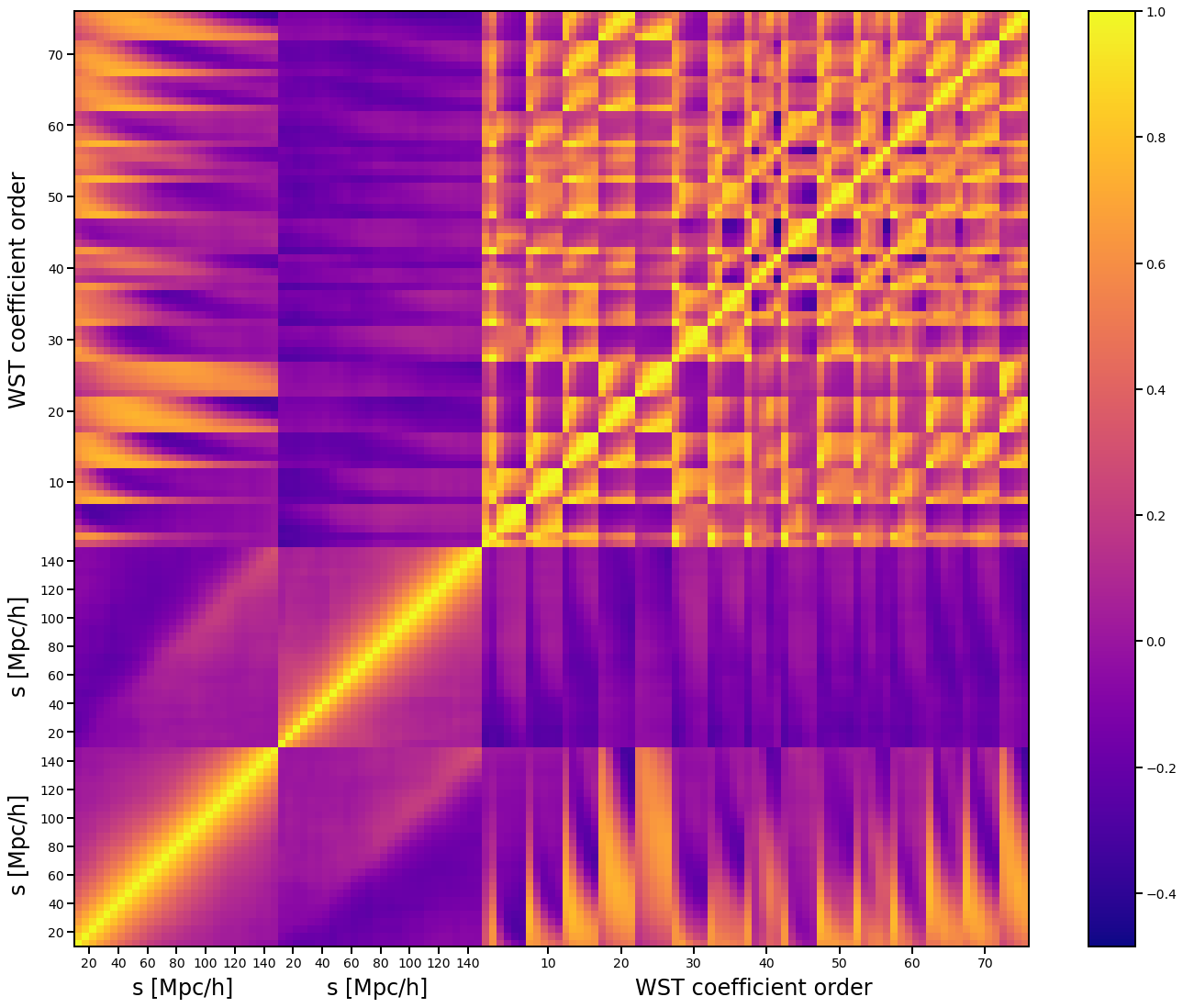}
\caption{Correlation matrix of the joint data vector consisting of the multipoles of the 2-point correlation function, $l=\{0,2\}$, and the 76 WST coefficients used in our analysis, evaluated from the 2048 realizations of the \textsc{Patchy} mocks for the fiducial cosmology. The lower left and upper right subplots coincide with the individual correlation matrices of the two statistics, respectively, while the rest corresponds to the cross-correlations between them. The WST coefficients populate the data vector in order of increasing values of the $j_1$ and $l_1$ indices, with the $l_1$ index varied faster. The $2\times2$ blocks on the lower left corner correspond to the auto- and cross- correlations of $\xi_0$ and $\xi_2$, from bottom to top and from left to right, respectively.}
\label{fig:covjoint} 
\end{figure*}

Furthermore, upon inversion of the covariance matrix in Eq. \eqref{eq:LogL}, we apply the standard de-biasing Hartlap correction factor \citep{refId22}:
\begin{equation}\label{Hartlap}
\hat{C}^{-1}= \frac{N_{\rm mocks}-N_{\bold{d}}-2}{N_{\rm mocks}-1}C^{-1}, 
\end{equation}
where $N_{\bold{d}}$ is the dimensionality of $\bold{X}_\bold{d}$, which will be $N_{\bold{d}}=76$ for the WST coefficients, $N_{\bold{d}}=24$ when working with the $l={0,2}$ multipoles of the correlation function (down to $r_{\rm min}=10.5$ Mpc/h), and $N_{\bold{d}}=100$ for the joint analysis. Before inverting, we make sure that the covariance matrices for both estimators are well-conditioned and can thus be safely inverted in order to be used in the likelihood in Eq. \eqref{eq:LogL}, and also that the number of realizations is sufficient for them to be well-converged (a very similar test for this can be found in Appendix B of Ref. \citep{PhysRevD.106.103509}). The correlation matrix, $C_{ij}/(C_{ii}C_{jj})$, of the joint statistic consisting of the 2-point function multipoles and the WST coefficients is shown in Fig.~\ref{fig:covjoint}, evaluated at the fiducial cosmology. Focusing on the WST coefficients on the upper right subplot, and starting with the $1^{st}$ order group of wavelets (that is, until index 25), we notice the existence of strong correlations between nearby scales and angles (close to the diagonal), which progressively decrease and turn into anti-correlations between the smallest and the largest wavelet scales. Similar patterns permeate into the $2^{nd}$ order group of wavelets and their correlations with the corresponding $1^{st}$ order scales. The correlation matrix of the 2-point correlation function multipoles, corresponding to the lower left corner, exhibits the familiar structure known in the literature \citep{10.1093/mnras/stv2826}. Lastly, when looking into the joint covariance between the two statistics in Fig.~\ref{fig:covjoint}, we observe the existence of positive correlations between the wavelets and the 2-point function monopole, which are most pronounced with the wavelets probing the largest scales. There are no significant correlations with the quadrupole of the correlation function, on the other hand. 

The final missing piece needed to evaluate the likelihood \eqref{eq:LogL} for a given point in the target parameter space is the theoretical dependence of the summary statistic as a function of the $8+7=15$ cosmological+HOD parameters, $\bold{X}_t(\theta)$, which we model using the emulator we trained (as explained in \S\ref{subsec:Emu}), which allows us to obtain accurate predictions in a fraction of a second.

Combining all of the above ingredients into our model, we sample the likelihood from Eq. \eqref{eq:LogL} using the Markov Chain Monte Carlo (MCMC) sampler {\tt emcee} \citep{2013PASP..125..306F}\footnote{Publicly available at \url{https://emcee.readthedocs.io/en/stable/}.}, so as to perform the posterior analysis of the CMASS dataset. 
Even though our original forward model spans a 15-dimensional parameter space, as explained in \S\ref{sec:simulation} and \S\ref{subsec:model}, our main focus is to obtain constraints on $\Lambda$CDM, so we fix $w_0=-1,w_a=0,a_{\rm run}=0, N_{\rm eff}=3.046$ (i.e. to their $\Lambda$CDM values) and define our baseline analysis to constrain the following $4+7=11$ cosmological+HOD $\Lambda$CDM parameters: $\theta=\{\omega_b, \omega_c, \sigma_8, n_s,\log M_{\rm cut},\log M_1, \sigma, \kappa, \alpha, \alpha_{\rm c},\alpha_{\rm s} \} $. We also obtain constraints on extensions to $\Lambda$CDM, for which our analysis will constrain the full 15-d parameter space consisting of 8 cosmological parameters, $\theta=\{\omega_b, \omega_c, \sigma_8, n_s, w_0,w_a,a_{\rm run}, N_{\rm eff}\} $ + the same 7 HOD nuissance parameters as above. We use flat priors bounded by the limits of the \textsc{AbacusSummit} simulations and the HOD training set, both of which are showed in Table~\ref{tab:hodbounds}.  
For parameter $\omega_b$, our baseline run is actually performed with a Gaussian prior:
\begin{equation}\label{omegabprior}
\omega_b = 0.02268 \pm 0.00038,
\end{equation}
as determined from Big Bang Nucleosynthesis (BBN) measurements, which is a choice commonly adopted in analyses of BOSS data \citep{Ivanov_2020,d_Amico_2020,Philcox_2020,Chen_2022}.
Finally, to confirm the sufficient convergence of our chains, we make sure that the mean value of the acceptance fraction falls within the reasonable range of values, $0.2-0.5$, and that the mean integrated auto-correlation time is at least 2 orders of magnitude lower than the total number of steps used, as suggested in Ref. \citep{2013PASP..125..306F}. Lastly, we make use of 8000 walkers, which are initialized in a tight ball around the {\it Planck} 2018 values.

%%%%%%%%%%%%%%%%%%%%%%%%%%%%%%%%%%%%%%%%%%%%%%
\subsubsection{\textsc{Patchy} mocks}\label{subsec:Patchy}

The covariance matrix of an estimator can be usually evaluated either using an analytical model under the Gaussian approximation or using simulations performed for multiple realizations at a given cosmology, through Eq. \eqref{eq:covmat}. Simulation-based analyses typically take advantage of covariance mocks, such as the $2048$ realizations of the publicly available \textsc{Multidark-Patchy} mocks\footnote{Available at \url{https://data.sdss.org/sas/dr12/boss/lss/dr12_multidark_patchy_mocks/}.} \citep{10.1093/mnras/stv2826,10.1093/mnras/stw1014}, hereafter referred to as \textsc{Patchy} mocks. We will use this collection for our BOSS analysis. 

The main reference simulation for this run \citep{10.1093/mnras/stw248} evolved $3840^3$ dark matter particles on a cubic volume of side $2.5$ Gpc/h, using the code {\tt GADGET-2} \citep{2005Natur.435..629S}, and for a baseline cosmology described by $\{\Omega_b,\Omega_m,n_s,\sigma_8,h\}=\{0.0482,,0.307,0.961,0.829,0.6778\}$. It was subsequently combined with an approximate perturbation theory-based gravity solver in order to produce mocks for the gravitationally bound halos, which were identified using the Bound Density Maximum halo finder \citep{Klypin:1997sk}. Finally, the galaxy mocks were created by populating the halos, using the Halo Abundance Matching technique \citep{Kravtsov_2004} in order to model the galaxy-halo connection. The \textsc{Patchy} mocks were also cut into the realistic survey geometry of BOSS CMASS, for both the NGC and SGC observed parts of the sky, while the systematic effects can be captured through a set of accompanied weights (in analogy to Eq. \eqref{eq:wc} for the data):
\begin{equation}\label{eq:wcPat}
w_{\rm c}(\bold{r}) = w_{\rm fc}(\bold{r})w_{\rm veto}(\bold{r}).
\end{equation}

Similar to Eq. \eqref{fkpfield}, the above weighting scheme captures fiber collisions, $w_{\rm fc}$, and the rest of the associated shortcomings of the dataset through a veto mask, $w_{\rm veto}$, while the FKP weights are also assigned through the usual Eq. \eqref{eq:fkpweight}. To evaluate the summary statistics from this set of mocks, we repeat the procedure detailed in \S\ref{subsec:summary_statistic} for Eq. \eqref{fkpfield}, but with the weighting scheme in Eq. \eqref{eq:wcPat}, as opposed to the one of Eq. \eqref{eq:wc} that we used for the data. For this purpose, the \textsc{Patchy} mocks are also accompanied by their own set of randoms containing $\sim 50 \times$ the number of objects in the corresponding actual galaxy mock.

Furthermore, we follow the standard procedure of assuming a cosmology-independent covariance matrix \citep{Kodwani_2019,refId0Car}, and convert the galaxy coordinates of the mocks, RA, DEC, and z, into comoving Cartesian ones using the fiducial cosmology of our forward model, which is the \textsc{c000} defined in \S\ref{sec:simulation}. As we also noted in Ref. \citep{PhysRevD.106.103509}, mixing different ways of modeling the cosmological dependence of the estimator and its covariance matrix is common practice in BOSS analyses (as in, e.g., Refs. \citep{10.1093/mnras/stw1096,10.1093/mnras/stw3298,Ivanov_2020,d_Amico_2020,Philcox_2020,PhysRevD.105.043517}). We combine two different sets of mocks (\textsc{AbacusSummit} $\&$ \textsc{Patchy}) in order to build our final model for the likelihood. 

It should be pointed out that, even though the \textsc{Patchy} mocks were also partly tested for their accuracy in capturing the 3-point correlation function of CMASS \citep{10.1093/mnras/stv2826,10.1093/mnras/stw1014}, in addition to the 2-point function, they have not been tuned for novel summary statistics such as the WST. This fact, combined also with their approximate gravity solver and the assumption of the cosmology-independent covariance matrix may be sources of error in our analysis, that we are working to overcome with the next generation of galaxy mocks designed to match the requirements for DESI analysis (see such an example in Ref. \citep{2024arXiv240109523P}).

%%%%%%%%%%%%%%%%%%%%%%%%%%%%%%%%%%%%%%%%%%%%%%
\subsection{Validation}\label{sec:Abactests}

In the previous section we described the detailed steps to perform a likelihood analysis using our emulator for the cosmological dependence of the WST coefficients. Before proceeding to analyze the actual CMASS dataset, we first test our pipeline to ensure its accuracy in inferring (known) cosmological parameters from simulated  data vectors.

%%%%%%%%%%%%%%%%%%%%%%%%%%%%%%%%%%%%%%%%%%%%%%
\subsubsection{Abacus hold-out mock tests}
\begin{figure*}[!ht]
    \subfloat{%
      \includegraphics[width=0.49\textwidth]{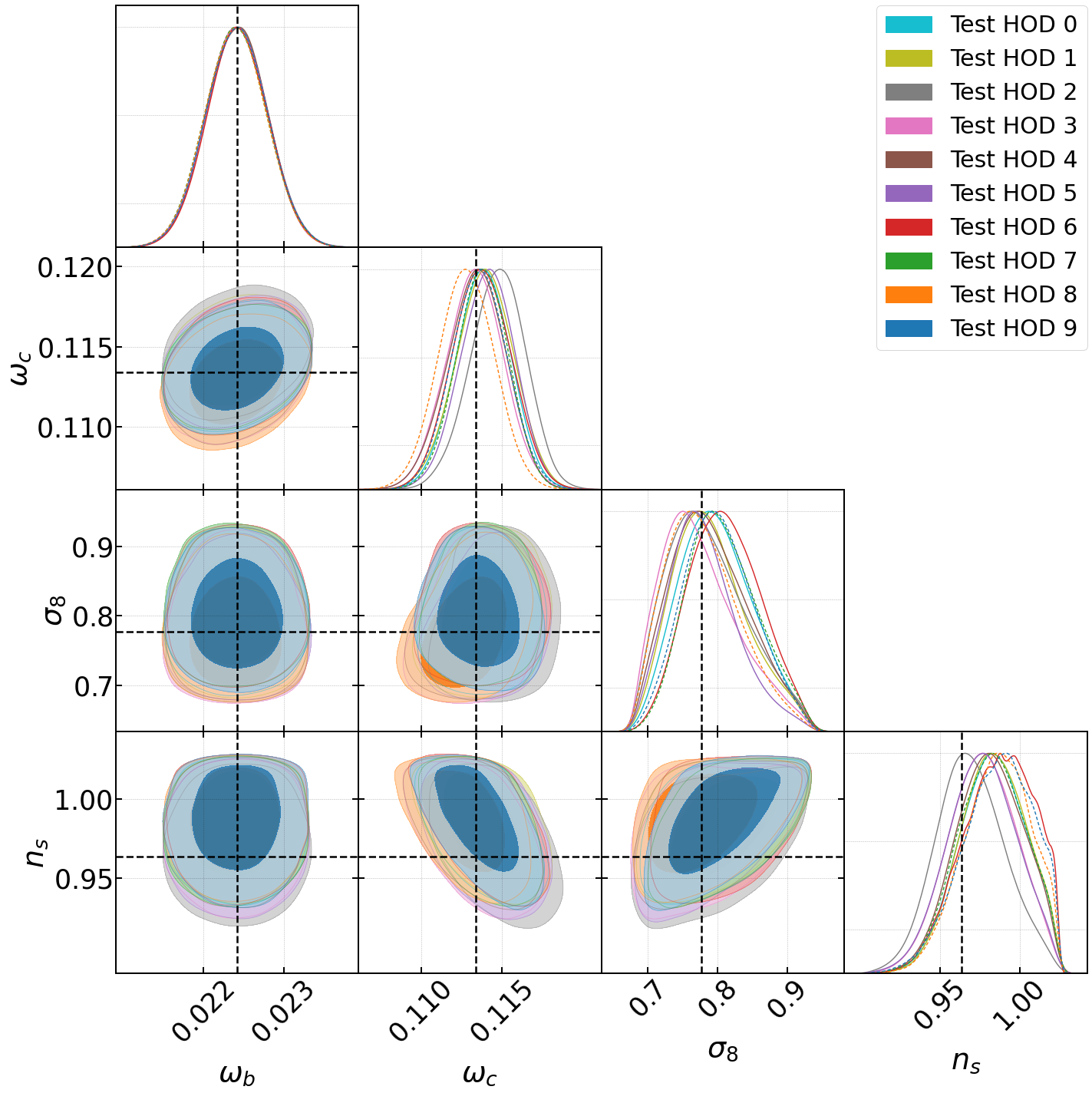}
    }
    \hfill
    \subfloat{%
      \includegraphics[width=0.49\textwidth]{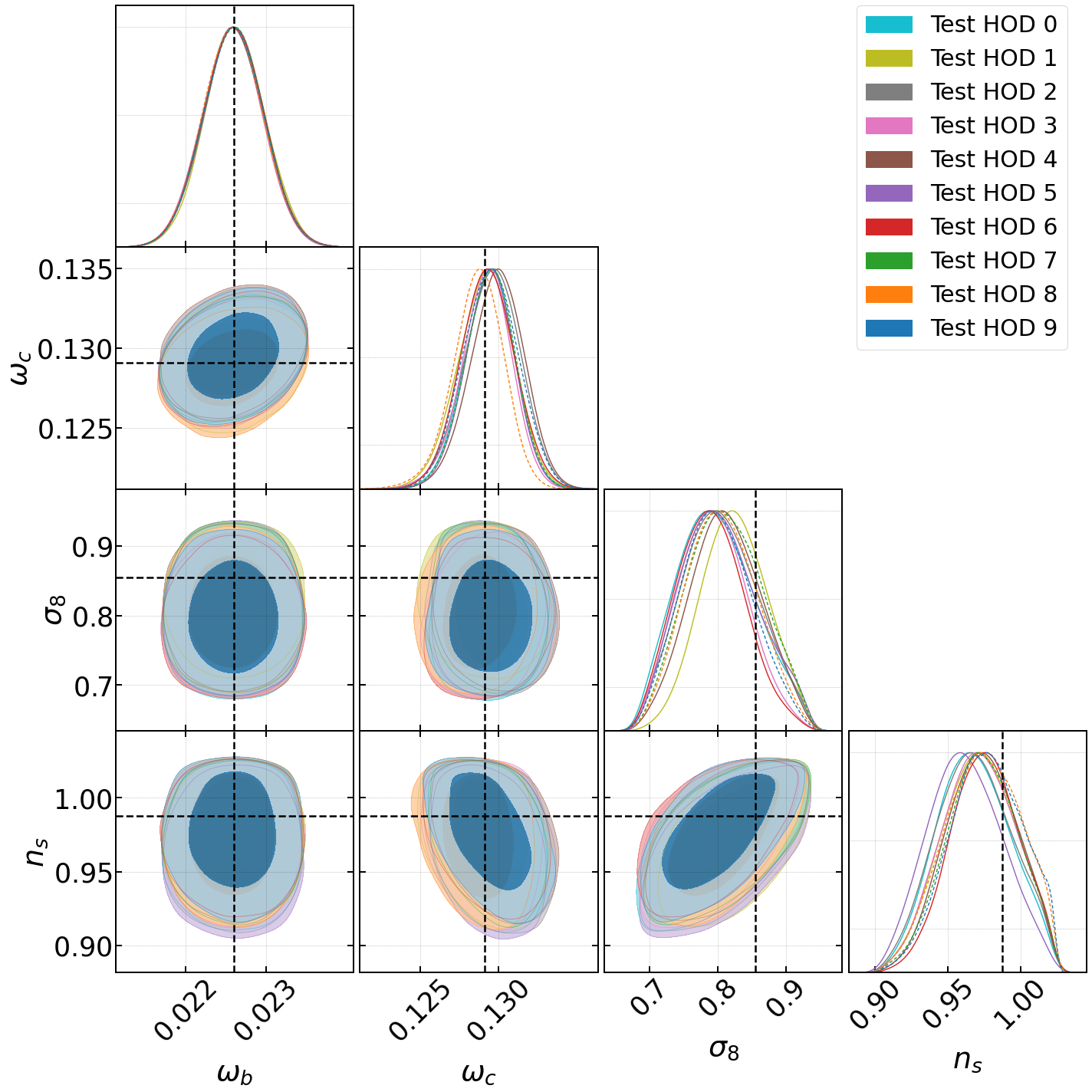}
    }
    \hfill
    \subfloat{%
      \includegraphics[width=0.49\textwidth]{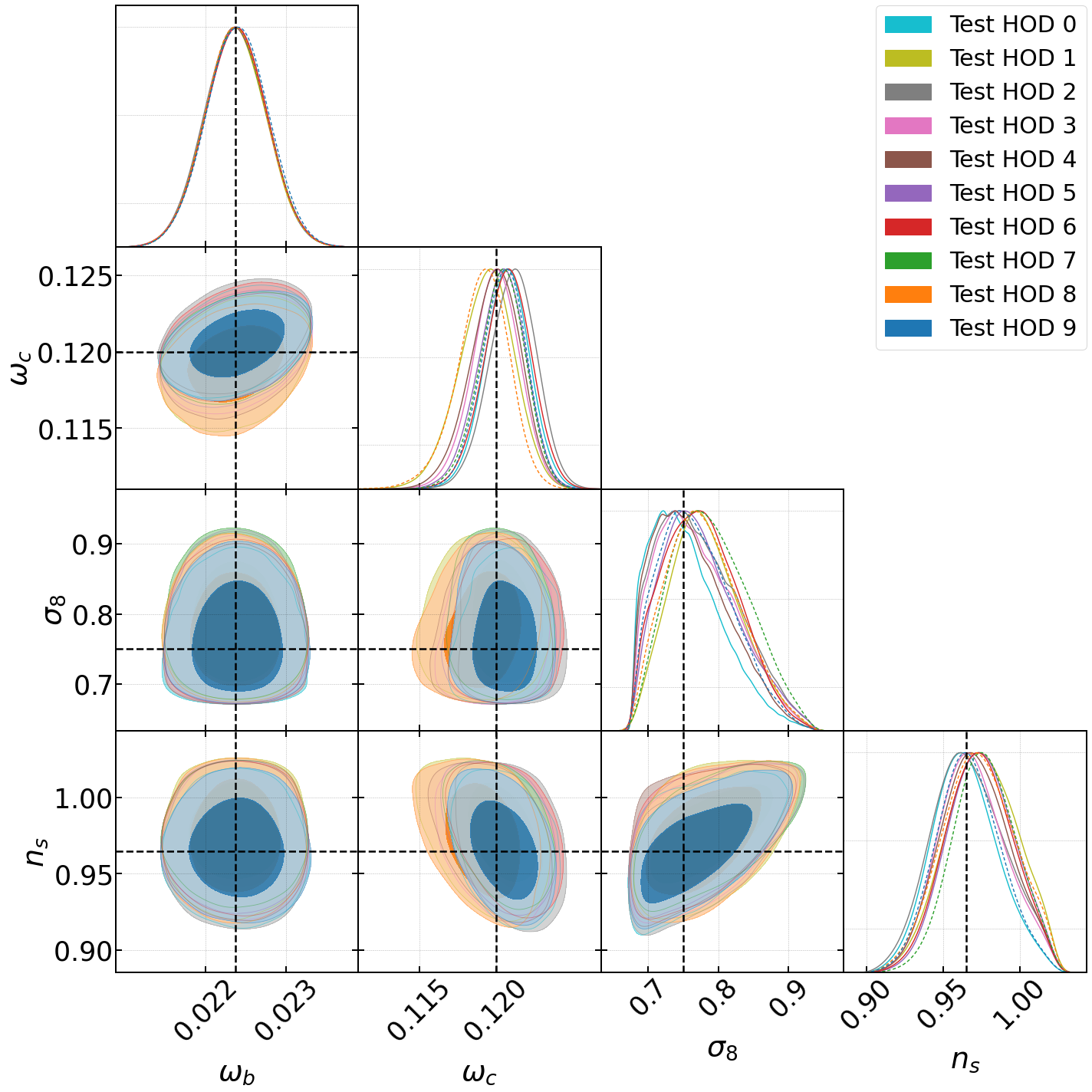}
    }
    \hfill
    \subfloat{%
      \includegraphics[width=0.49\textwidth]{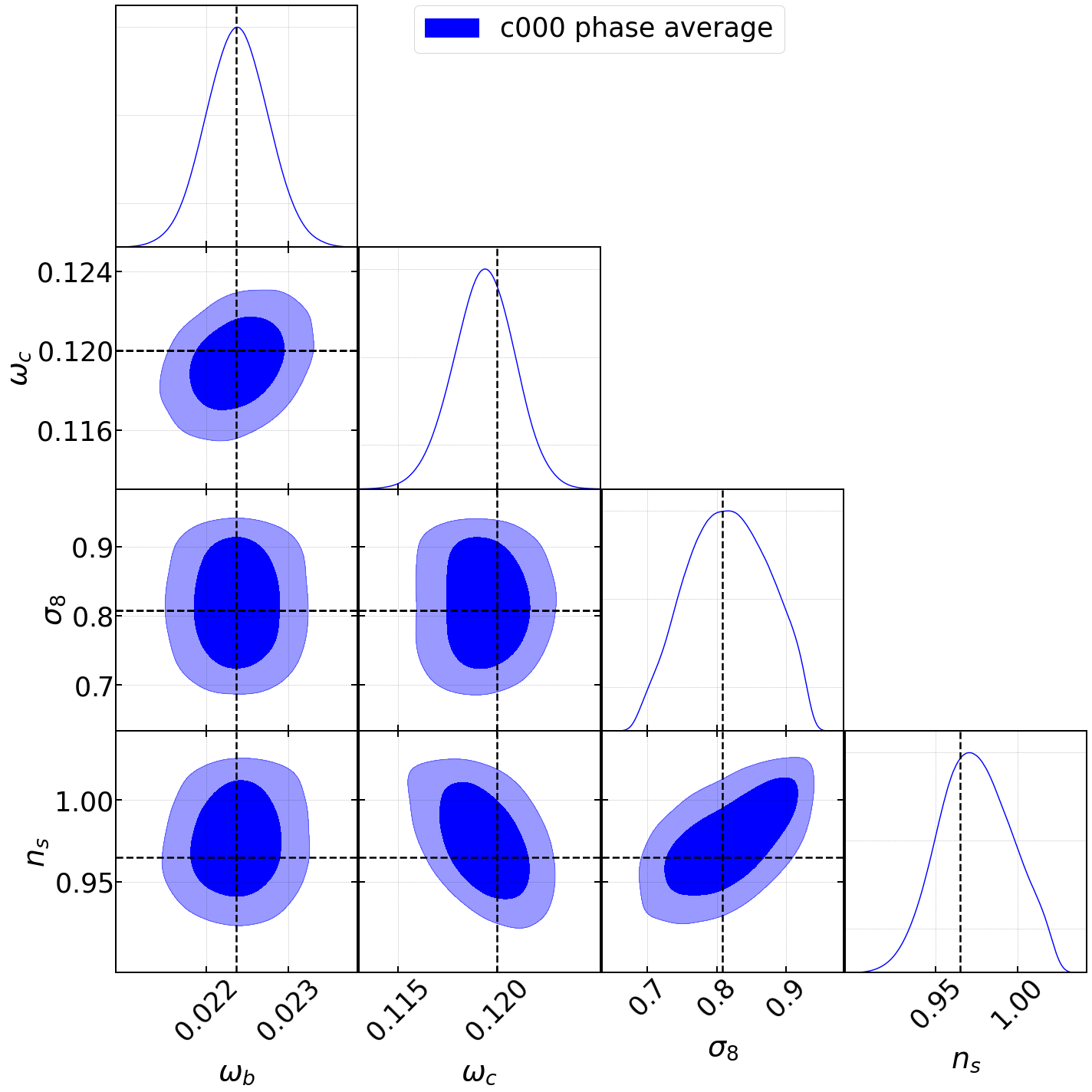}
    }
    \caption{$\Lambda$CDM recovery tests using our WST emulator to analyze 10 HOD configurations of the \textsc{c001} (upper left), \textsc{c003} (upper right) and \textsc{c004} (lower left) hold-out cosmologies of our test set. We also show the marginalized 1-$\sigma$ and 2-$\sigma$ posteriors obtained by analyzing the mean data vector of the 24 additional realizations available for the fiducial \textsc{c000} cosmology (lower right). The horizontal and vertical black dashed lines indicate the true values of the cosmological parameters in each case.}
    \label{fig:recovery}
  \end{figure*}

In order to test the accuracy of our inference pipeline, we begin by randomly selecting 10 HOD configurations centered around the best-fit values from Ref. \citep{2022MNRAS.515..871Y}, for each one of the hold-out \textsc{c001-004} \textsc{AbacusSummit} cosmologies. We repeat all previously explained steps to produce synthetic WST data vectors from each test mock, which are then fed into our likelihood analysis pipeline to constrain the cosmological parameters of our $\Lambda$CDM baseline case. The corresponding marginalized posterior distributions obtained on the 4 $\Lambda$CDM cosmological parameters of the baseline analysis are then shown in Fig.~\ref{fig:recovery}, in which we see that we are able to recover the true values within 1-$\sigma$ levels of accuracy, for all cases. We note that we do not show the contours for the wCDM test cosmology \textsc{c002} in Fig.~\ref{fig:recovery}, for brevity, but the recovery is successful in this case as well.

The hold-out cosmologies used for the above tests correspond to the same initial fixed phase, ph000, of the \textsc{AbacusSummit} simulations as the mocks of our training set, and as a result do not allow us to detect potential biases introduced by this approximation. To check for this, we also attempt to perform parameter inference from the data vector obtained by averaging over the 24 additional phases, ph001-024, that are available for the fiducial \textsc{c000} cosmology. As we also show in Fig.~\ref{fig:recovery}, our phase correction scheme \eqref{eq:phasecorr} is found sufficient to recover an accurate cosmology from a different phase. We add that we confirmed the recovery was also successful when we used these 24 phases individually, as the mock data vector. 

Overall, the above tests confirm that our WST emulator, in combination with the error correction strategies \eqref{eq:covmatall} $\&$ \eqref{eq:phasecorr}, is successful in inferring the parameters of the \textsc{AbacusSummit} simulations within 1-$\sigma$ levels of accuracy, over a wide range of cosmologies. We should also note, at this point, that we confirmed the same to be true for the corresponding emulator for the multipoles of the galaxy correlation function. 

%%%%%%%%%%%%%%%%%%%%%%%%%%%%%%%%%%%%%%%%%%%%%%
\subsubsection{\textsc{Uchuu} mock tests}
Even though our simulation-based model was found to be successful at recovering the true cosmological parameters over a broad range of tests, as reported in the previous section \S\ref{sec:Abactests}, the corresponding hold-out mocks that we used were produced from the same set of the \textsc{AbacusSummit} simulations, using, more importantly, the same assumptions for the galaxy-halo connection through the specific HOD model we adopted (described in \S\ref{subsec:model}). As a result, before our pipeline can be trusted for a reliable interpretation of the actual observations, it should be first tested against an independent simulation with different gravity and halo codes, with a different strategy for the population of dark matter halos with galaxies. 

To achieve this goal, we additionally make use of the Uchuu simulations \citep{2021MNRAS.506.4210I,2022arXiv220800540D,2022arXiv220714689O, 2023MNRAS.519.1648A, Prada2023:2304.11911}, which were run using the GreeM N-body code \citep{Ishiyama2009:0910.0121}. It evolved 2.1$\times 10^{12}$ dark matter particles, in a simulation volume $(2 \ \rm Gpc/h)^3$, which matches the one of Abacus, and it is large enough to fit the entire footprint of BOSS. Their underlying cosmology corresponds to the following values: $\Omega_m = 0.3089$, $\Omega_b = 0.0486$, $h = 0.6774$, $\sigma_8 = 0.8159$, and $n_s = 0.9667$, while dark matter halos were identified with the \textsc{Rockstar} halo finder \citep{2010ApJ...717..379B}, in contrast to \textsc{Abacus}'s {\sc CompaSO}.

More crucially, the corresponding galaxy mocks were produced with \textsc{UniverseMachine}~\citep[UM;][]{2019MNRAS.488.3143B}, a model that is considerably more sophisticated than the HOD. UM is an empirical galaxy--halo connection model that predicts galaxy star formation rates from halo mass and halo assembly histories. It is a flexible framework that models the full evolution histories of galaxies anchored on dark matter halo merger trees from cosmological simulations, and it is simultaneously constrained by observed galaxy stellar mass functions, UV luminosity functions, quenched fractions, cosmic star formation history, and galaxy clustering over a wide range of galaxy mass and redshifts (up to $z\sim 8$). We refer the readers to \cite{2023MNRAS.519.1648A} for detailed descriptions of the mock. It is also worth highlighting that UM naturally includes a motivated yet flexible prescription of galaxy assembly bias, as the galaxy properties are directly computed from the halo merger trees. Thus, this test also checks against potential systematic biases due to galaxy assembly bias. 

For the covariance matrix needed for the \textsc{Uchuu} likelihood analysis we use the same suite of \textsc{Patchy} mocks described in \S\ref{subsec:Patchy}, since both types of simulations are tuned to match the clustering properties of the BOSS CMASS sample, with a same volume and number density and a similar Planck-like cosmology.

\begin{figure}[h!]
\centering 
\includegraphics[width=0.5\textwidth]{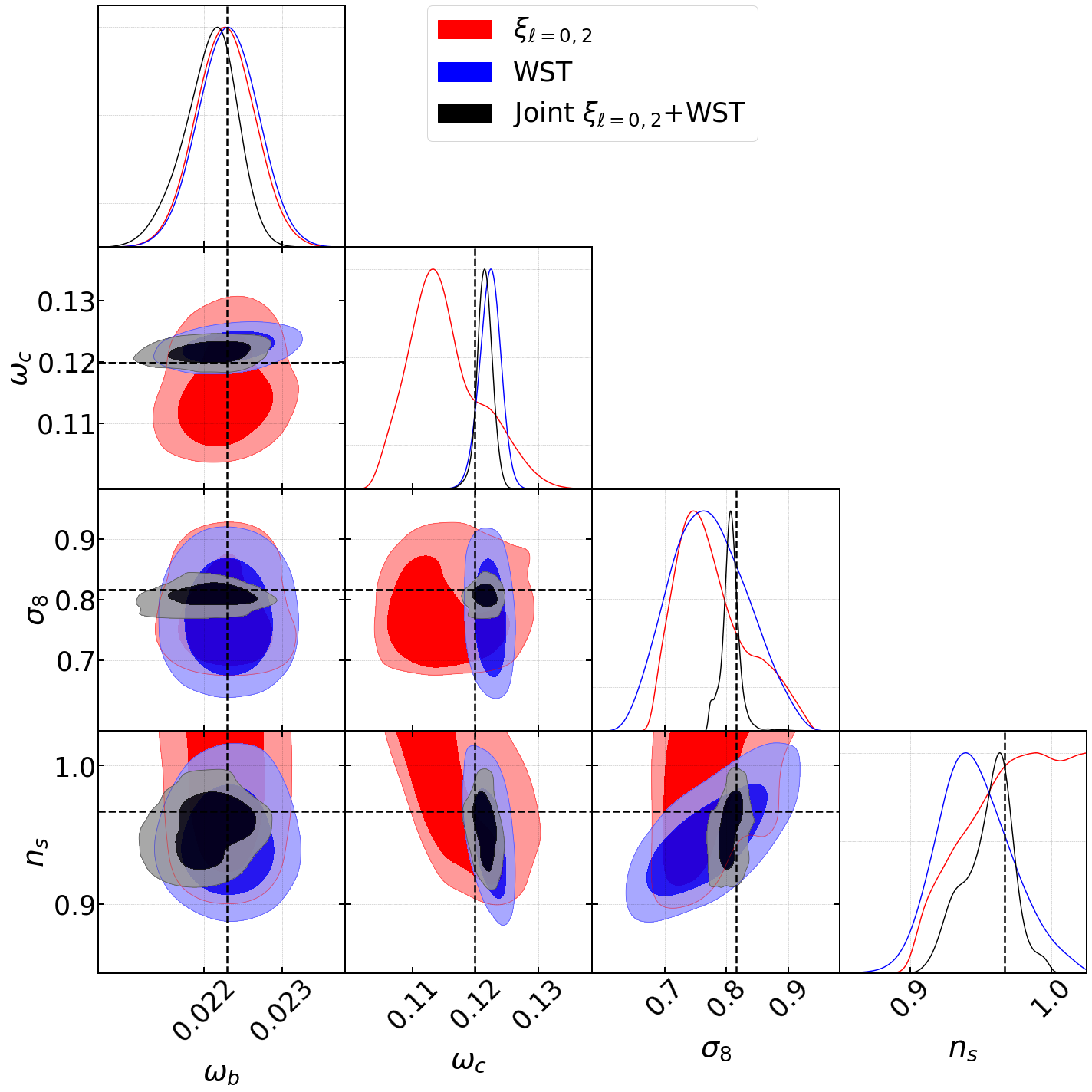}
\caption{Recovery test using the Uchuu galaxy mock for the $\Lambda$CDM cosmological parameters obtained using the monopole and quadrupole of the galaxy correlation function (red), the WST coefficients (blue) and their joint combination (black). The horizontal and vertical black dashed lines indicate the true values of the cosmological parameters.}
\label{Fig:Uchuu}
\end{figure}

In Fig.~\ref{Fig:Uchuu}, we plot the marginalized constraints obtained on the 4 $\Lambda$CDM parameters after analyzing the Uchuu mock using the multipoles of the galaxy correlation function, the WST coefficients and a joint combination of both. As in the previous case of the \textsc{Abacus} hold-out tests, we find that the true values always lie within 1-$\sigma$ away from the mean, for all 3 cases. We note that $n_s$ is prior-dominated in the case of the 2-point correlation function, as the contour hits the upper prior bound of the \textsc{AbacusSummit} grid\footnote{A similar finding was recently reported by Ref. \citep{Cuesta-Lazaro2023:2309.16539}.}. This is not the case, however, for the WST and the joint combination, which are the main focus of this analysis, despite the significantly tighter 1-$\sigma$ errors they predict. These results confirm our ability to trust that our forward model can recover unbiased cosmological constraints which are robust against the various assumptions made by the simulations used for its training.

%%%%%%%%%%%%%%%%%%%%%%%%%%%%%%%%%%%%%%%%%%%%%%
\section{Results}\label{sec:Results}

{
\begin{table*}
\centering

\setlength{\extrarowheight}{2.6pt}
\begin{tabular}{|p{0.49cm}||p{3.55cm}||p{3.55cm}||p{3.55cm}|}
\hline 
& 2-point c.f. & WST & Joint 2-point c.f.+WST\\
\end{tabular}

\centering
\begin{tabular}{|p{0.5cm}||p{1.2cm}|p{2.2cm}||p{1.2cm}|p{2.2cm}||p{1.2cm}|p{2.2cm}|}

\hline 
 & Best-fit& Mean$\pm \sigma$& Best-fit& Mean$\pm \sigma$& Best-fit& Mean$\pm \sigma$\\
\hline 
$\omega_b$ & $0.02261$&$0.02270^ {+0.00037}_{-0.00037}$&$0.02274$&$0.02277^{+0.00038}_{-0.00038}$&$0.0225$&$0.02262^{+0.00029}_{-0.00029}$\\
\hline
$\omega_c$ &$0.1201$&$0.1222^{+0.0040}_{-0.0063}$&$0.1239$&$0.1244^{+0.0015}_{-0.0015}$&$0.1238$&$0.1241^{+0.0011}_{-0.0011}$\\
\hline
$n_s$ &$0.925$&$0.922^{+0.037}_{-0.037}$&$0.961$&$0.951^{+0.023}_{-0.023}$&$0.927$&$0.924^{+0.01}_{-0.01}$\\
\hline
$\sigma_8$ &$0.742$&$0.746^{+0.051}_{-0.051}$&$0.860$&$0.834^{+0.058}_{-0.039}$&$0.793$&$0.795^{+0.019}_{-0.019}$\\
\specialrule{.2em}{.1em}{.1em}
$h$ & $0.677$&$ 0.677^{+0.022}_{-0.015}$&$0.67$&$0.669^{+0.0059}_{-0.0059}$&$0.668$&$0.669^{+0.0049}_{-0.0049}$\\
\hline
\end{tabular}

\caption{Best-fit values, mean values and $68\%$ confidence intervals for all cosmological parameters resulting from the likelihood analysis of the 2-point correlation function multipoles (left), the WST coefficients (middle) and a joint analysis of the two (right). The mean values are presented in the format `${\rm mean}^{+1 \sigma}_{-1 \sigma}$', after marginalization over all HOD parameters.}
\label{table:2}
\end{table*}
}

\begin{figure*}[ht!]
\centering 
\includegraphics[width=0.89\textwidth]{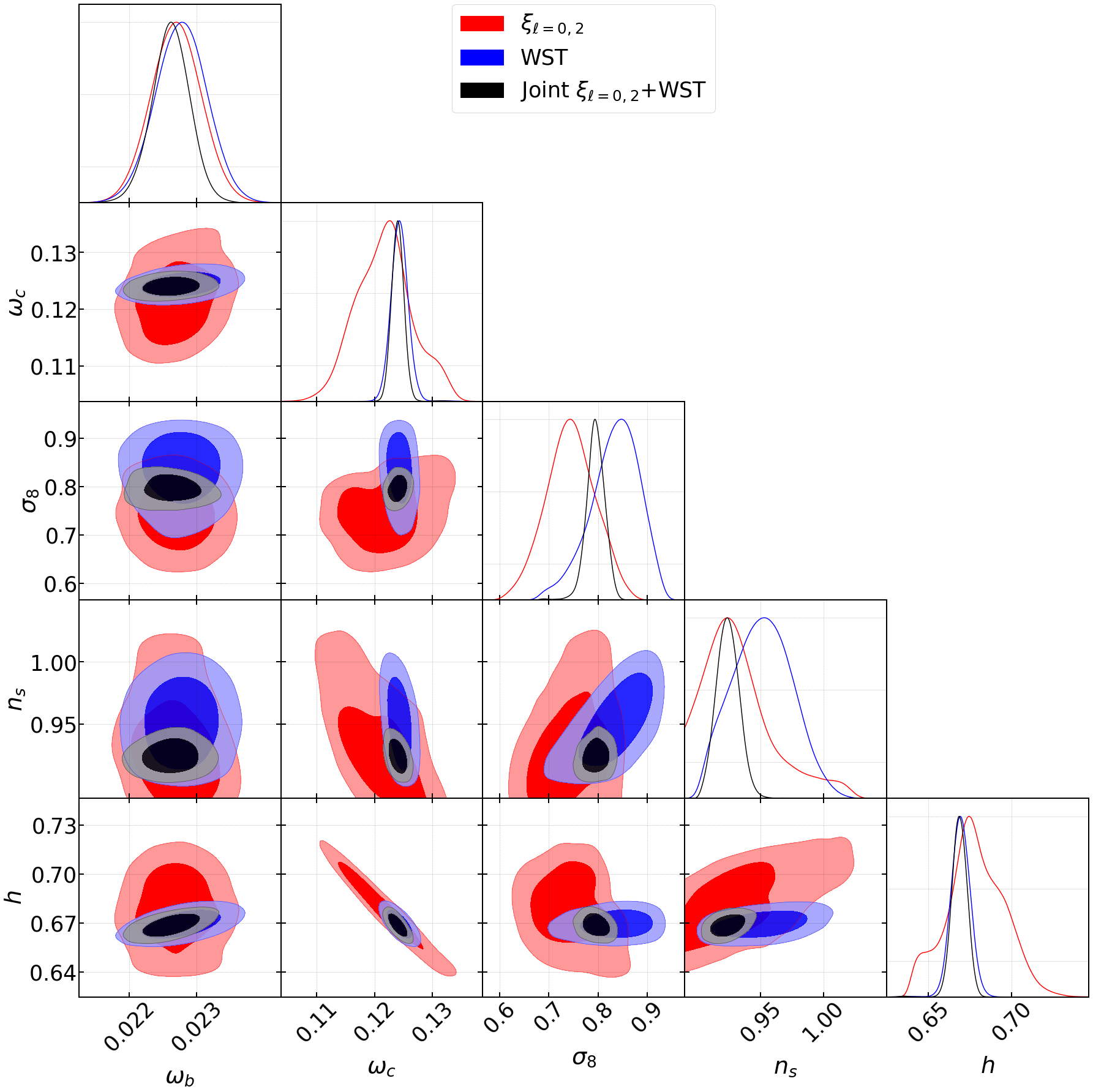}
\caption{Marginalized constraints on the $\Lambda$CDM cosmological parameters obtained using the monopole and quadrupole of the galaxy correlation function (red), the WST coefficients (blue) and their joint combination (black) in order to analyze the BOSS CMASS observations. The results shown above were obtained after imposing a BBN Gaussian prior on the value of $\omega_b = 0.02268 \pm 0.00038$.}
\label{Fig:Corner5params}
\end{figure*}

Having validated our pipeline against a series of internal and external mock recovery tests, described in \S\ref{sec:Abactests}, we now proceed to use it in order to analyze the BOSS CMASS dataset. Specifically, in Fig.~\ref{Fig:Corner5params} we plot the 2-dimensional marginalized posterior probability distributions of the 4 $\Lambda$CDM parameters of our baseline CMASS analysis, as they were obtained using the multipoles of the galaxy 2-point correlation function, the WST coefficients and their joint combination. We also show the constraints on the dimensionless Hubble constant, $h$, that we obtain by treating it as a \textit{derived} parameter from our MCMC chains, resulting from the fixed value of the comoving angular size of the sound horizon at last scattering, $100 \theta_{\star}=1.041533$, imposed in the \textsc{AbacusSummit} simulations. We note that, even though this parameter is very well-constrained by the {\it Planck} satellite \citep{Planck2018}, this choice implies that $h$ is not varied independently in our inference, so the corresponding result should be interpreted with caution. In addition, the mean and $1\sigma$ error values obtained on the cosmological parameters (marginalized over HOD) are listed in Table~\ref{table:2}, while the corresponding constraints on the HOD nuisance parameters are presented in Appendix \S\ref{Appsec:HOD}.

We begin with the standard analysis using the galaxy correlation function, the results of which are broadly consistent with {\it Planck} 2018 \citep{Planck2018}. Even though the mean values obtained for $n_s$ and $\sigma_8$ are somewhat lower than the ones of {\it Planck}, the magnitude of these differences is not statistically significant ($\sim 1\sigma$), unlike the results of some previous BOSS analyses (Eg. \citep{2022MNRAS.515..871Y,PhysRevD.105.043517,Chen_2022}).

Moving on to discuss the results of the WST re-analysis, we first notice the relative consistency between the corresponding mean values for the parameters extracted from the two estimators, the differences of which never exceed the respective $1-\sigma$ values obtained from the correlation function. We do, however, notice different degeneracy directions exhibited by the WST contours projected on the various individual 2-d parameter planes, the importance of which will become apparent below. More importantly, the 1-$\sigma$ errors obtained on parameters $\omega_c$ and $n_s$ are found to be 4.2 and 1.6$\times$ tighter than the corresponding predictions from the correlation function, as seen in Table~\ref{table:2}. The dimensionless Hubble constant is consistent with Planck in this case as well, with an error that is 3.7$\times$ tighter than from the correlation function, tracing the respective results for $\omega_c$, on which it depends through the fixed $\theta_{\star}$. We note again that this finding should be interpreted with caution. if it were not for the strong prior on $\theta_{\star}$ in our model, the coarse logarithmic binning used by our wavelets would likely not be able to fully capture the BAO information, resulting in a less accurate determination of $h$. Last but not least, we do not find any noticeable improvement (with respect to the 2-point function) in our ability to constrain $\sigma_8$, while the mean value predicted by the WST analysis is also consistent with Planck. Even though counter-intuitive, at face value, this result is attributed to the inclusion of the residual emulator error, $C_{\rm emu}$, in Eq. \eqref{eq:covmatall}, a fact that we have tested and confirmed, as shown in Appendix \S\ref{Appsec:Cemu}. In particular, we find that, even though the intrinsic errors predicted by the WST in the limit of zero emulation error are substantially tighter for all parameters, our actual error budget is also much larger for the WST, such that the addition of $C_{\rm emu}$ in Eq. \eqref{eq:covmatall} partially or completely (in the case of $\sigma_8$) masks the net improvements. We nevertheless choose to include this term, in order to ensure a reliable and robust analysis. 

We have already seen that, despite the fact that the WST and correlation function contours are consistent with each other at the 1$\sigma$ level, they exhibit different degeneracy orientations. This is not as surprising if we consider that the two statistics do not capture the exact same information. Indeed, we remind that, even at the lowest ($1^{st}$) order, the WST raises the modulus of the input galaxy density field to the power $q=0.8$, closer to the properties of other higher-order statistics such as the marked power spectrum, as we also found in Ref.  \citep{PhysRevD.105.103534}. The localized solid harmonic wavelet \eqref{solid:sup} is also different than the Fourier kernel of the power spectrum/correlation function, with all the additional known benefits associated with this choice \citep{Sifre_2013_CVPR,10.1214/14-AOS1276,6822556,cheng:2021xdw}. As a consequence, analyzing the data with the joint combination of the two statistics allows us to break degeneracies and further improve upon the results obtained from each individual analysis, as we can see in Fig.~\ref{Fig:Corner5params} and Table~\ref{table:2}. In particular, the 1$\sigma$ error obtained on $\sigma_8$, which previously did not improve by a WST application alone, now shrinks by a factor of 2.5, while the corresponding constraints on the rest of the parameters are further tightened by a factor of $3-6$ compared to the 2-point correlation function and by a factor of $1.4-2.5$ compared to the WST-only results. Overall, the joint analysis allows us to constrain the parameters $\omega_c$, $\sigma_8$, $n_s$, and $h$ with $0.9\%$, $2.3\%$, $1\%$ and $0.7\%$ levels of accuracy, respectively. This result, which can be considered to be the main one of our work, highlights the value held in a complementary analysis employing both the WST coefficients and the standard correlation function.

\begin{figure*}[b]
\centering 
\includegraphics[width=0.89\textwidth]{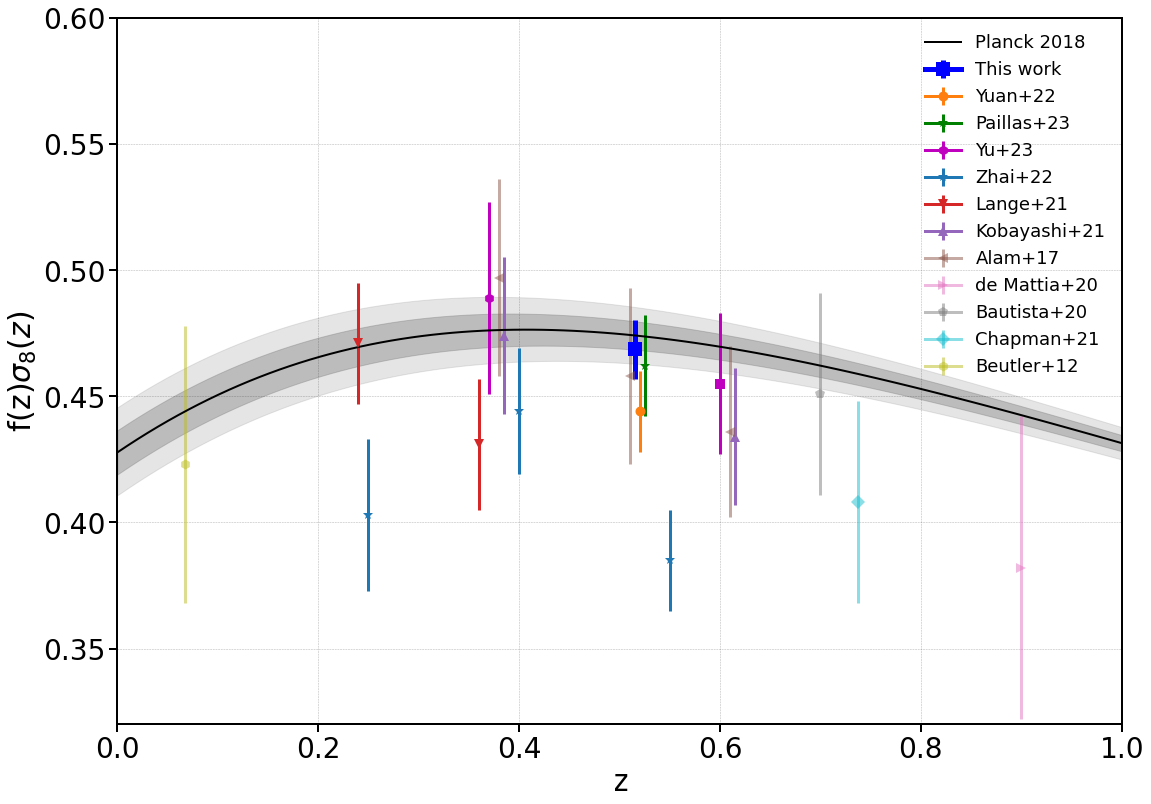}
\caption{Marginalized constraints on the structure growth rate, $f(z)\sigma_8 (z)$, of our joint analysis in blue alongside other clustering constraints in the literature. We show the {\it Planck} 2018 \citep{Planck2018} CMB constraints in black, with the corresponding 68$\%$ and 95$\%$ limits in shaded bands, together with the results from two other recent Abacus-based CMASS re-analyses using the small-scale 2-point correlation function \citep{2022MNRAS.515..871Y} and the density split clustering statistic \citep{Paillas2023:2309.16541,Cuesta-Lazaro2023:2309.16539}. Additionally, we show clustering constraints from BOSS LOWZ small-scale RSD \citep{2022MNRAS.509.1779L}, BOSS full-shape power spectrum \citep{2021Kobayashi}, BOSS large-scale RSD+BAO \citep{BOSS:2016wmc}, BOSS small-scale RSD \citep{2023Zhai} eBOSS small-scale RSD \citep{Chapman:2021hqe}, eBOSS large-scale RSD+BAO \citep{Bautista:2020ahg,deMattia:2020fkb}, BOSS DR12 large-scale power spectrum \citep{Yu_2023} and the 6dF Galaxy Survey \citep{10.1111/j.1365-2966.2012.21136.x}.}
\label{Fig:fsigma8}
\end{figure*}

In addition to the above parameters, and in order to align our analysis with a standard practice adopted by many RSD studies in the literature, we further quote results on the product $f(z)\sigma_8(z)$, with $f(z)=\frac{d \ln D(a)}{da}$ and $D(a)$ being the linear growth rate and growth factor, respectively. This is also a derived parameter that we obtain from the samples in our chains. In particular, at the effective redshift, $z_{\rm eff}=0.515$, of our sample, the joint analysis gives: 
\begin{equation}\label{fsig}
f\sigma_8(z_{\rm eff}=0.515) = 0.469 \pm 0.012,
\end{equation}
which corresponds to a determination at a $2.5 \%$ level of accuracy. Furthermore, in Fig.~\ref{Fig:fsigma8} we plot our result together with the corresponding one from {\it Planck} 2018 \citep{Planck2018} and from a selected sample of other recent BOSS re-analyses in the literature (which we will further discuss shortly). Our prediction is consistent with {\it Planck} well within the 1$\sigma$ levels, driven by the corresponding consistency in our inferred value of $\sigma_8$, and despite our relatively higher value of $\omega_c$. In the context of lensing studies, this can be alternatively examined in terms of the parameter combination $S_8 = \sqrt{\left(\Omega_m / 0.30 \right)}\sigma_8$, for which we get 
\begin{equation}\label{eq:S8}
S_8  = 0.833 \pm 0.023,
\end{equation}
in almost perfect agreement with the fiducial {\it Planck} result, $S_8  = 0.832 \pm 0.013$. 

As far as the values of $\omega_c$ and $n_s$ are concerned, they are found to be statistically higher and lower, respectively, relative to the Planck result, driven by the very tight constraints of our joint analysis. It is interesting to notice that a similar trend has been found in some recent large-scale BOSS analyses, when $n_s$ is left free \citep{PhysRevD.105.043517}. The magnitude of the tension was smaller in these studies, however, due to the larger error bars produced by such perturbation theory-based models. Ref.~\citep{2022MNRAS.515..871Y} also found a preference for a lower $n_s$ at the 1.5$\sigma$ level.

We also note that, even though all above results were produced assuming a tight BBN prior \eqref{omegabprior} on $\omega_b$, we found that our joint analysis is actually able to constrain this value reasonably well even with a flat, uninformative prior, as shown in Appendix \S\ref{Appsec:omegabpriors}. 

\begin{figure}[ht]
\includegraphics[width=0.48\textwidth]{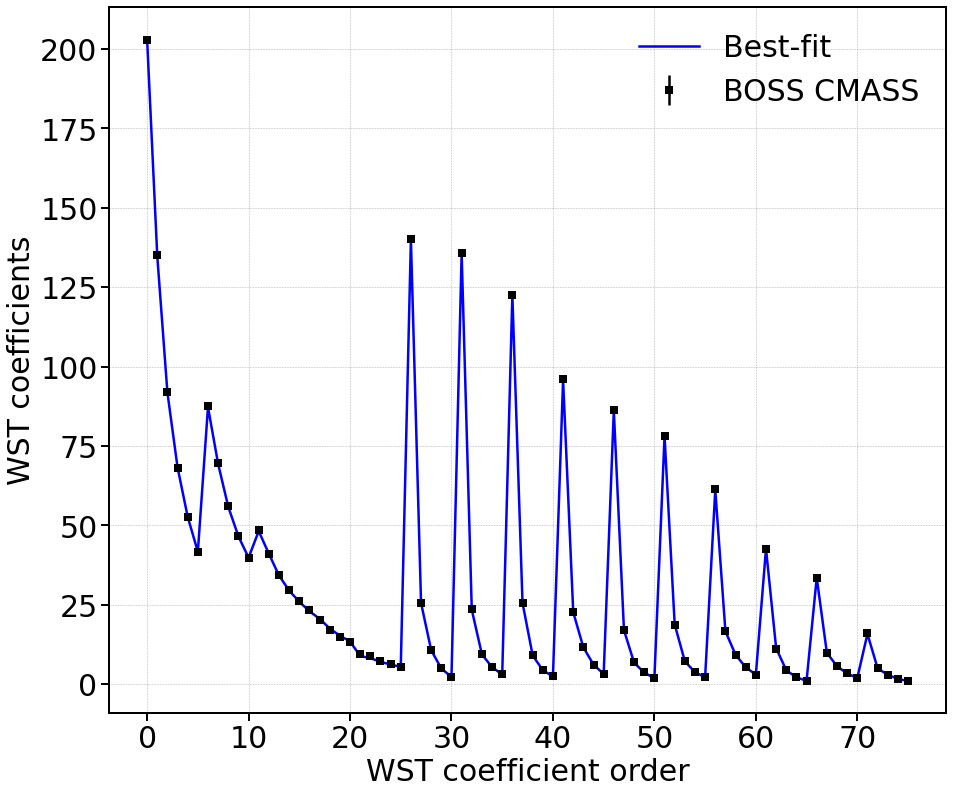}
\caption{All 76 WST coefficients evaluated from the BOSS CMASS dataset (black circles) plotted together with the best-fit prediction obtained from our likelihood analysis (solid blue line). The WST coefficients populate the data vector in order of increasing values of the $j_1$ and $l_1$ indices, with the $l_1$ index varied faster.}
\label{fig:WSTfit} 
\end{figure}

Furthermore, in the left sub-panels of Table~\ref{table:2} we list the best-fit values obtained for each one of the 3 types of analyses considered in this work, which are always found to lie within a standard deviation away from the corresponding means. To assess the goodness-of-fit, we also evaluate the $\chi^2$ per degrees of freedom (d.o.f.), $\chi_{\nu}^2\equiv \chi^2 /$d.o.f., which is found to be equal to 1.11, 1.36 and 1.37 for the correlation function, the WST and the joint analysis, respectively. The result for the 2-point function is very similar to the one reported by the Abacus-based small-scale CMASS analysis of \citep{2022MNRAS.515..871Y}. Even though the value for the WST is a bit higher, it is still reasonable and within the same range and/or lower than the corresponding results reported by recent analyses using other higher-order statistics, such as k-nearest neighbors \citep{yuan2023robust} and density-split statistics \citep{Paillas2023:2309.16541}. The goodness of the fit is also visually evident in Fig.~\ref{fig:WSTfit}, in which we plot the best-fit prediction for the WST together with the corresponding CMASS measurement.

We should comment, at this point, on how our new results compare with the ones of our previous WST BOSS analysis \citep{PhysRevD.106.103509}, which relied on a simple Taylor expansion approximation to model the cosmological dependence of the WST coefficients. Starting with the 1-$\sigma$ errors, the main difference for parameters $\sigma_8$ and $n_s$ is caused by the inclusion of the WST emulator error, $C_{\rm emu}$, in Eq. \eqref{eq:covmatall}, as we also pointed out above and show in Appendix \S\ref{Appsec:Cemu}. In fact, we note that, if we omit this contribution, the WST errors on $\sigma_8$ and $n_s$ become 0.027 and 0.02, respectively, which are not that far off from what we previously reported, as seen in Table II of Ref.~\citep{PhysRevD.106.103509}. Even after neglecting the emulator error, the constraint on $\omega_c$ was still found to be $\sim 2.5\times$ tighter in Ref. \citep{PhysRevD.106.103509}, a fact that is most likely attributed to the simplified Taylor expansion approximation. This fact is also most likely responsible for the relatively low value of $\sigma_8$ that we reported in that work.

Finally, we briefly discuss how our work compares against other analyses of BOSS clustering in the literature, starting with the two other recent applications of the \textsc{AbacusSummit}. Ref. \citep{2022MNRAS.515..871Y} built an Abacus-based emulator of the anisotropic 2-d correlation function to analyze the CMASS dataset, while Refs. \citep{Paillas2023:2309.16541,Cuesta-Lazaro2023:2309.16539} worked with the density-split clustering statistic. Given that they relied on the same suite of simulations for their forward model, these applications share the same cosmology grid and priors as our analysis (including the fixed value of sound horizon $\theta_{*}$). All three analyses also used the same HOD framework. However, there are several key differences between the above works and ours, that need to be pointed out: given the unique sensitivity of the WST to the survey geometry, through the successive wavelet convolutions in Eq. \eqref{eq:WSTcoeff:base}, we trained our emulator using the cut-sky mocks matching the exact CMASS footprint, rather than the original cubic boxes used by \citep{2022MNRAS.515..871Y,Paillas2023:2309.16541,Cuesta-Lazaro2023:2309.16539}. For similar reasons, we worked with a flattened density profile, $n(z)$, and in a slightly narrower redshift cut, as we showed in Fig.~\ref{fig:CMASSnz}. Furthermore, both of the other works included smaller scales down to 1 $\mathrm{Mpc / h}$ and thus accounted for the necessary effects of assembly bias in their HOD parametrization, which we neglected given that our analysis stopped at a minimum scale of $\sim10 \ \mathrm{Mpc / h}$. Ref. \citep{2022MNRAS.515..871Y} also used Jackknife re-sampling in order to compute the covariance matrix (as opposed to our \textsc{Patchy} mocks) and included only small scales, < $30 \ \mathrm{Mpc / h}$, in their analysis, while Ref.~\citep{Paillas2023:2309.16541} only analyzed the NGC part of the BOSS survey. Due to all the above differences, a direct ``apples to apples'' comparison is still hard. Nevertheless, we notice the relative 1$\sigma$ consistency between our results for $\sigma_8$ and $f \sigma_8$ and those of \citep{Paillas2023:2309.16541} (and Planck), as also seen in Fig.~\ref{Fig:fsigma8}. The analysis of Ref. \citep{2022MNRAS.515..871Y}, on the other hand, found a relatively lower value of the clustering amplitude, which, combined with their tighter errors, leads to a disagreement at the level of $1.5 \sigma$. In addition to the previously mentioned differences, other analysis choices that might be driving this difference are the use of Gaussian priors by Ref. \citep{2022MNRAS.515..871Y} and the fact that their emulator error was evaluated drawing from from the posterior, rather than the prior.  In preparation for the analysis of the next stage of spectroscopic observations, we plan to revisit these comparisons through a commonly adopted set of uniform analysis choices. 

A plethora of other studies in the literature have analyzed the BOSS and extended BOSS (eBOSS) \citep{Dawson:2015wdb} observations, including, but not limited to, the ones plotted in Fig.~\ref{Fig:fsigma8} alongside our result. All of them used the standard 2-point correlation function or the power spectrum, and can be grouped into large-scale \citep{BOSS:2016wmc,Bautista:2020ahg,deMattia:2020fkb,Yu_2023,10.1111/j.1365-2966.2012.21136.x} and small-scale studies \citep{2021Kobayashi,2022MNRAS.509.1779L,2023Zhai,Chapman:2021hqe}. Despite the large variance in the modeling and analysis choices among the members of this list, we notice that our analysis joins the ones that are statistically consistent with the Planck curve, including the official BOSS result \citep{BOSS:2016wmc}. On the other hand, a number of studies is found to systematically underpredict the growth rate, related to the known LSS tension that has emerged in the last few years. Given that the true origin of this discrepancy is not yet known, we hope that novel techniques such as the WST will help shed light on this issue. We highlight that our constraint is found to be the tightest reported among all these studies. Similar considerations apply for the comparison to other BOSS analyses, e.g., 
\citep{Ivanov_2020,d_Amico_2020,Philcox_2020,PhysRevD.105.043517,Chen_2022,Zhang_2022,2022arXiv220410392C}.

{
\begin{table}
\centering

\setlength{\extrarowheight}{2.6pt}
\begin{tabular}{|p{1.1cm}||p{3.55cm}|}
\hline 
& Joint 2-point c.f.+WST\\
\end{tabular}

\centering
\begin{tabular}{|p{1.1cm}||p{1.2cm}|p{2.2cm}|}

\hline 
 & Best-fit& Mean$\pm \sigma$\\
\hline 
$\omega_b$ &$0.02280$& $0.02273^{+0.00036}_{-0.00036}$\\
\hline
$\omega_c $ &0.1227&$0.1239^{+0.0056 }_{-0.0056}$\\
\hline
$\sigma_{8}  $ &$0.748$&$0.751^{+0.034}_{-0.040}$\\
\hline
$n_s  $ &$0.928$&$0.953^{+0.022}_{-0.030} $\\
\specialrule{.2em}{.1em}{.1em}
$h  $ &$0.675$&$0.671^{+0.021}_{-0.021 }   $\\
\specialrule{.2em}{.1em}{.1em}
$a_{\rm run}  $ &$0.002$&$0.004^{+0.019}_{-0.012}  $\\
\hline
$N_{\rm eff}  $ &$3.048$&$3.23^{+0.26}_{-0.26}$\\
\hline
$w_0  $ &$-1.039$&$-0.995^{+0.061}_{-0.073} $\\
\hline
$w_a  $ &$0.29$&$0.17^{+0.24}_{-0.21}   $\\
\hline
\end{tabular}

\caption{Best-fit values, mean values and $68\%$ confidence intervals for all cosmological parameters resulting from the joint WST + correlation function likelihood analysis in the case of the extended cosmological scenario. The mean values are presented in the format `${\rm mean}^{+1 \sigma}_{-1 \sigma}$', after marginalization over all HOD parameters.}
\label{Table:beyond}
\end{table}
}

\begin{figure*}[b]
\centering 
\includegraphics[width=0.89\textwidth]{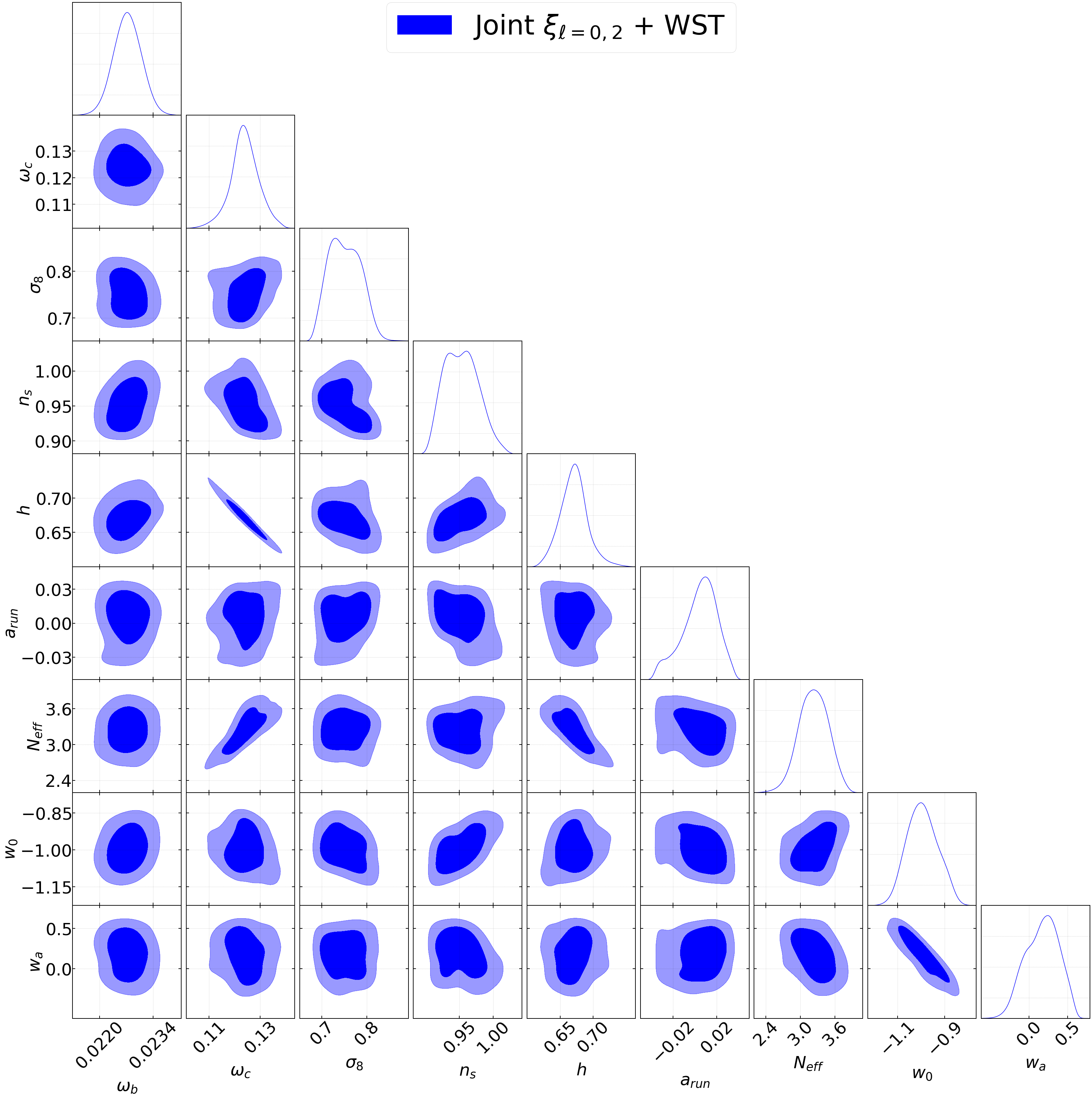}
\caption{Marginalized constraints on extensions to $\Lambda$CDM obtained using the joint WST + correlation function combination in order to analyze the BOSS CMASS observations. The results shown above were obtained after imposing a BBN Gaussian prior on the value of $\omega_b = 0.02268 \pm 0.00038$.}
\label{Fig:Cornerbeyond}
\end{figure*}

\subsection{Constraints on $\Lambda$CDM extensions}\label{subsec:beyond}
Our main focus for the present analysis has been to obtain constraints on $\Lambda$CDM. However, as we explained in \S\ref{sec:simulation}, our emulator was originally trained on the \textsc{AbacusSummit} cosmology grid, which also includes 4 additional parameters describing extensions to $\Lambda$CDM: $\{a_{\rm run}, N_{\rm eff}, w_0, wa\}$. As a result, and in order to further explore the constraining capabilities of the WST, here we briefly present constraints from the base joint WST + correlation function likelihood analysis on all 8 cosmological parameters (marginalized over the 7 HOD parameters), shown in Fig.~\ref{Fig:Cornerbeyond} and Table~\ref{Table:beyond}. We find that our analysis is able to clearly constrain all parameters simultaneously, without any signs of statistically significant deviations away from the known $\Lambda$CDM limits, $w_0=-1,w_a=0,a_{\rm run}=0, N_{\rm eff}=3.046$. Given the increased number of parameters in this case, it is not surprising that the constraints on the $\Lambda$CDM parameters are looser compared to the corresponding values found in the base analysis. As a consequence of the same fact, the previously reported tensions for $\omega_c$ and $n_s$ are alleviated in this case, and all $\Lambda$CDM parameters are found to be consistent with the \textit{Planck} 2018 results \citep{Planck2018} (and also with the ones of the base analysis). The reduced $\chi^2$/d.o.f is found to be $\chi_{\nu}^2=1.31$, confirming that the fit is equally good as the one of the joint base analysis.

We note that our pipeline has been more thoroughly tested for $\Lambda$CDM applications, and these results are exploratory in nature, while we reserve a more detailed WST application to extended scenarios for future work. Nevertheless, they serve as an additional example that showcases the promise held in the use of the WST in the context of parameter inference applications.

%%%%%%%%%%%%%%%%%%%%%%%%%%%%%%%%%%%%%%%%%%%%%%
\section{Conclusions}\label{sec:Conclusions}

In this work, we perform a thorough re-analysis of the BOSS CMASS DR12 dataset, using a simulation-based emulator for the Wavelet Scattering Transform, a novel statistic that promises to capture non-Gaussian information in a clustered field by subjecting it to a series of successive wavelet-convolutions.

In our series of previous works \citep{PhysRevD.105.103534,PhysRevD.106.103509}, we laid the foundation for a WST application to spectroscopic galaxy data, including the methodology to capture all necessary associated layers of realism to achieve this task, such as the effects of non-trivial survey geometry, the shortcomings of the dataset through a set of systematic weights or the Alcock-Paczynski effect. However, in order to reduce the related computational cost, in Ref. \citep{PhysRevD.106.103509} we used a linear Taylor expansion to approximate the cosmological dependence of the WST estimator. 

Having the full suite of the state-of-the-art \textsc{AbacusSummit} simulations at our disposal, we now revisit our previous analysis after constructing an accurate neural net-based emulator for the cosmological dependence of the WST coefficients. Our forward model is trained using a total of 151,474 mocks that span a 15-dimensional parameter space, capturing variations in 8 cosmological parameters and 7 Halo Occupation Distribution (HOD) nuissance parameters to model the galaxy-halo connection. We repeat these steps to create a corresponding emulator for the standard multipoles of the galaxy 2-point correlation function, which serves as our benchmark to evaluate the performance of the WST.

In order to ensure that our likelihood analysis pipeline achieves the necessary levels of accuracy for a reliable and robust cosmological analysis, we subject it to a series of internal and external parameter recovery tests. For the internal tests, we use 40 hold-out mocks that span a broad range of cosmological parameters within our training set. We then test and confirm that we can accurately infer the parameters of an external simulation, that used different assumptions to capture the complicated physics of galaxy formation. 

After confirming the accuracy of our forward model, we use it to re-analyze the BOSS CMASS DR12 dataset in the redshift range $0.4613<z<0.5692$, in order to constrain the $\Lambda$CDM parameters using the WST coefficients and the multipoles of the galaxy correlation function. We find that a joint analysis using the WST and the correlation function allows us to constrain the $\Lambda$CDM parameters with 1$\sigma$ errors that are tighter by a factor of $2.5-6$, compared to the 2-point correlation function, and by a factor of $1.4-2.5$ compared to the WST-only results. This corresponds to a competitive $0.9\%$, $2.3\%$, $1\%$ and $0.7\%$ level of determination for parameters $\omega_c$, $\sigma_8$, $n_s$, and $h$, respectively. Furthermore, the joint analysis allows us to obtain a tight $2.5 \%$ constraint on the parameter combination $f(z)\sigma_8(z)$, in agreement with the 2018 results of the \textit{Planck} satellite. We discuss how our new results compare against our previous analysis and prior ones in the literature, reaffirming the constraining power of the WST.

We also obtained constraints on extended cosmological scenarios, parametrized through 4 additional parameters, $\{a_{\rm run}, N_{\rm eff}, w_0, wa\}$, finding no statistically significant deviations from the $\Lambda$CDM limit.

Our emulator for the cosmological dependence of the WST coefficients (and the correlation function) has allowed us to overcome the main limitation behind our previous application \citep{PhysRevD.106.103509}. There is, however, room for further improvement in certain components of our forward model, which we plan to achieve in future work. First of all, and as we already pointed out above, the \textsc{AbacusSummit} simulations impose a fixed value of the angular scale $\theta_{\star}$. Even though this quantity is very well constrained by CMB observations \citep{Planck2018}, such a prior implies that the Hubble constant is not independently varied in our chains, but can only be obtained as a derived parameter. Given, however, that our framework is flexible enough to be applied to any set of simulated mocks, this limitation can be easily overcome using a different training set. In a similar manner, our use of simulations produced at a fixed redshift, $z=0.5$, implies that clustering evolution along the CMASS lightcone is currently neglected. With the capability to produce \textsc{Abacus} lightcones already in place \citep{2023Yuan,Hadzhiyska:2023fic}, we plan to incorporate this effect in future revisions of our pipeline. Furthermore, our current HOD parametrization for the galaxy-halo connection neglected assembly bias, given the more conservative scale-cut we adopted. It would be very interesting, in future work, to explore the full small-scale constraining power of the WST, for which a more general HOD model including assembly bias would be necessary. Such an endeavor will also require a more careful treatment of systematic effects, such as fiber collisions, which we currently corrected using the recipe designed for the standard correlation function analysis (for an example, see Ref. \citep{2023Yuan}).

The culmination of this series of works opens up an avenue of potentially exciting cosmological applications of the WST, with the advent of the first Stage-IV spectroscopic observations by DESI. As we had also pointed out in Ref. \citep{PhysRevD.106.103509}, the basis of solid harmonic wavelets that we have been using is not optimized for a spectroscopic dataset, as it was designed in the context of isotropic 3-d applications of molecular chemistry. A suitably tailored new basis of wavelets could potentially fully leverage the anisotropic RSD information in the observed galaxy field, by treating a given direction as special \citep{2021arXiv210411244S}. Higher-order statistics, as we have discussed before \citep{PhysRevD.97.023535,PhysRevD.105.103534}, also exhibit tremendous potential for constraining fundamental physics such as massive neutrinos, theories of gravity or primordial non-Gaussianity, through their unique ability to break degeneracies that are present at the power spectrum level. All of these are very interesting avenues that we would like to explore, alongside the first WST application to the first year of DESI data.

Our application serves as a prime example of how novel estimators, such as the Wavelet Scattering Transform, can hopefully allow us to fully exploit the vast amount of information that will be accessed by the next generation of cosmological surveys, giving us the opportunity to potentially revolutionize our fundamental understanding of the universe.

%%%%%%%%%%%%%%%%%%%%%%%%%%%%%%%%%%%%%%%%%%%%%%
\vspace{0.5cm}
\subsection*{Acknowledgments} 
We would like to thank Carolina Cuesta-Lazaro, Daniel Eisenstein, Hector Gil-Marin, Johannes Lange, Enrique Paillas and Peter Taylor for useful discussions over the course of this work.

GV acknowledges the support of the Eric and Wendy Schmidt AI in Science Postdoctoral Fellowship at the University of Chicago, a Schmidt Futures program. This work is supported by the National Science Foundation under Cooperative Agreement PHY-
2019786 (The NSF AI Institute for Artificial Intelligence and Fundamental Interactions, http://iaifi.org/). CD and GV have also been partially supported by the Department of Energy (DOE) Grant No. DE-SC0020223.

The massive production of all MultiDark-Patchy mocks for the BOSS Final Data Release has been performed at the BSC Marenostrum supercomputer, the Hydra cluster at the Instituto de Fısica Teorica UAM/CSIC, and NERSC at the Lawrence Berkeley National Laboratory. That work acknowledges support from the Spanish MICINNs Consolider-Ingenio 2010 Programme under grant MultiDark CSD2009-00064, MINECO Centro de Excelencia Severo Ochoa Programme under grant SEV- 2012-0249, and grant AYA2014-60641-C2-1-P. The MultiDark-Patchy mocks was an effort led from the IFT UAM-CSIC by F. Prada’s group (C.-H. Chuang, S. Rodriguez-Torres and C. Scoccola) in collaboration with C. Zhao (Tsinghua U.), F.-S. Kitaura (AIP), A. Klypin (NMSU), G. Yepes (UAM), and the BOSS galaxy clustering working group.

We thank Instituto de Astrofisica de Andalucia (IAA-CSIC), Centro de Supercomputacion de Galicia (CESGA) and the Spanish academic and research network (RedIRIS) in Spain for hosting Uchuu DR1 and DR2 in the Skies \& Universes site for cosmological simulations. The Uchuu simulations were carried out on Aterui II supercomputer at Center for Computational Astrophysics, CfCA, of National Astronomical Observatory of Japan, and the K computer at the RIKEN Advanced Institute for Computational Science. The Uchuu DR1 and DR2 effort has made use of the skunIAA RedIRIS and skun6IAA computer facilities managed by the IAA-CSIC in Spain (MICINN EU-Feder grant EQC2018-004366-P).

\newpage

%%%%%%%%%%%%%%%%%%%%%%%%%%%%%%%%%%%%%%%%%%%%%%
\appendix

%%%%%%%%%%%%%%%%%%%%%%%%%%%%%%%%%%%%%%%%%%%%%%%%%%%%
\section{Gaussianity of the WST likelihood}\label{Appsec:Gaussianity}

Our posterior analysis in \S\ref{sec:Analysis} has been performed by sampling from a likelihood that we assumed to follow a Gaussian form, as given by Eq. \eqref{eq:LogL}. In the standard power spectrum case, and despite the non-Gaussianity of the cosmic  density field at late times, this is known to be an accurate approximation thanks to the Central Limit Theorem, when a sufficiently large number of modes contribute to the value evaluated at a given spatial bin. For the WST, in our previous applications \citep{PhysRevD.105.103534,PhysRevD.106.103509} we also adopted this approximation, motivated by supportive findings in the 2D weak-lensing (WL) applications of Ref. \citep{10.1093/mnras/stab2102}. We now proceed to explicitly test and confirm the validity of this assumption for the present WST application to 3D clustering, using the $2048$ realizations of the \textsc{Patchy} mocks for the fiducial cosmology. Following Refs. \citep{2021MNRAS.508.3125F,2022arXiv220904310P}, the $2048$ realizations will have a $\chi^2$ distribution, given by:
\begin{equation}\label{chisq}
\chi^2_i = \left[\bold{X}_\bold{d_i}-\bar{\bold{X}}_{\bold{d}}\right]^{\rm T} C^{-1}\left[\bold{X}_\bold{d_i}-\bar{\bold{X}}_{\bold{d}}\right],
\end{equation}
where $\bold{X}_{\bold{d_i}}$ is the prediction for the $i^{th}$ \textsc{Patchy} mock realization, $\bar{\bold{X}}_{\bold{d}}$ the mean value over the distribution, and $C$ the covariance matrix from Eq. \eqref{eq:covmat}.

\begin{figure}[ht!]
\centering 
\includegraphics[width=0.49\textwidth]{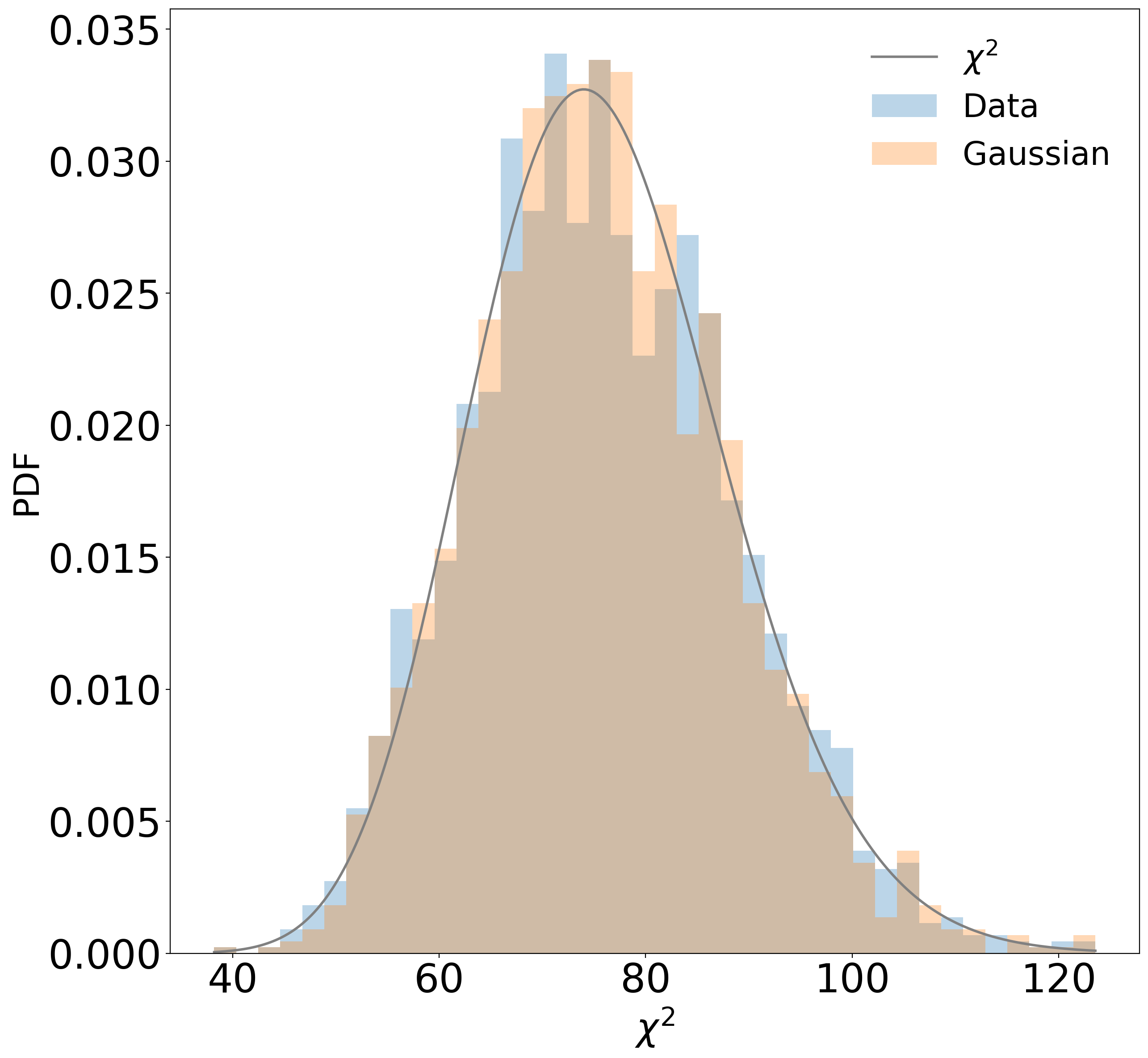}
\caption{Probability density function of the $\chi^2$ distribution of the WST coefficients as measured from the $2048$ realizations of the \textsc{Patchy} mocks (blue) plotted together with a theoretical $\chi^2$ distribution with $N_{\bold{d}}=76$ degrees of freedom (black line) and a Gaussian distribution with the same mean and covariance (orange). The WST estimator does not exhibit any significant deviations from a Gaussian distribution.}
\label{Fig:Gausstest}
\end{figure}
\begin{figure}[ht!]
\centering 
\includegraphics[width=0.49\textwidth]{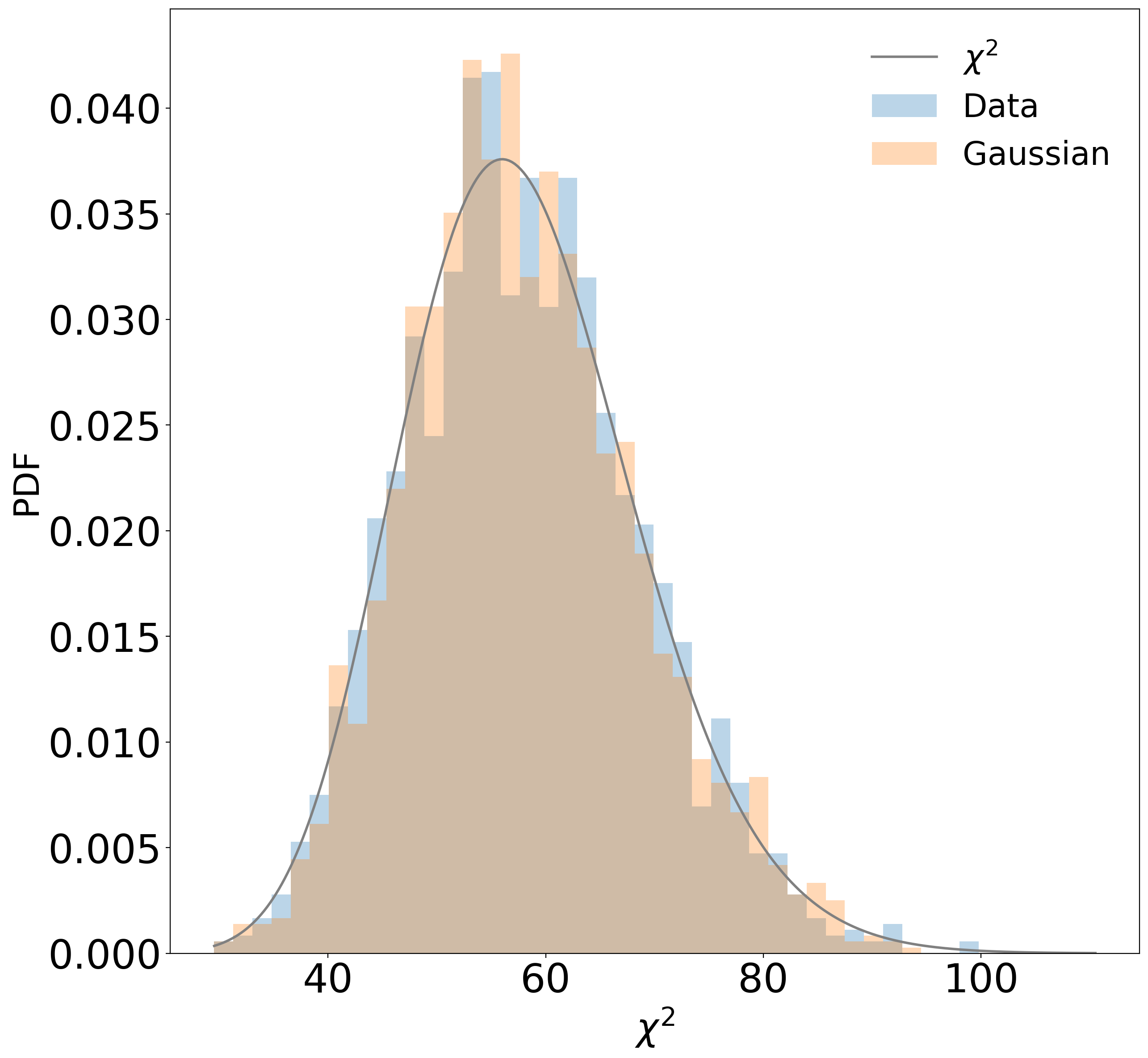}
\caption{The same $\chi^2$ analysis as in Fig. \ref{Fig:Gausstest} is repeated for the multipoles of the 2-point correlation function that we use as the benchmark in our analysis.}
\label{Fig:Gausstestxi}
\end{figure}
If the likelihood of a summary statistic is indeed Gaussian, then the probability density function (pdf) from Eq. \eqref{chisq} should closely track the theoretical $\chi^2$ distribution with degrees of freedom equal to the dimensionality of the data vector (ie. $N_{\bold{d}}=76$ for our WST implementation). It should also closely match the pdf of samples randomly drawn from a Gaussian distribution with the same mean and covariance as the sample of realizations. This comparison is demonstrated in Fig.~\ref{Fig:Gausstest} for the WST, which is observed to satisfy a high level of consistency between the 3 curves, confirming thus a high degree of Gaussianity for the likelihood of the WST estimator. The equivalent comparison for the 2-point correlation function multipoles is shown in Fig.~\ref{Fig:Gausstestxi} for reference, which reproduces the known result of Gaussianity of the correlation function.
This result for the Gaussianity of the WST is aligned with the one of Ref. \citep{10.1093/mnras/stab2102} in the context of weak lensing and also with the results of Refs. \citep{2021MNRAS.508.3125F,2022arXiv220904310P,2023Yuan} for other higher-order statistics explored in the literature. 

We also note that a quantification of the Gaussianity of various summary statistics was performed in Ref. \cite{Park:2022hzj}, in which the probability distribution of the WST coefficients evaluated from the simulated 3D matter density field was found to exhibit a certain degree of non-Gaussianity. However, this work used a different basis of wavelets, that performed a much finer sampling of the spatial domain, which can lead to a breakdown of the Central Limit Theorem. As a result, their findings are not inconsistent with ours.

%%%%%%%%%%%%%%%%%%%%%%%%%%%%%%%%%%%%%%%%%%%%%%%%%%%%
\section{Constraints on HOD parameters}\label{Appsec:HOD}

\begin{figure*}[ht!]
\centering 
\includegraphics[width=0.99\textwidth]{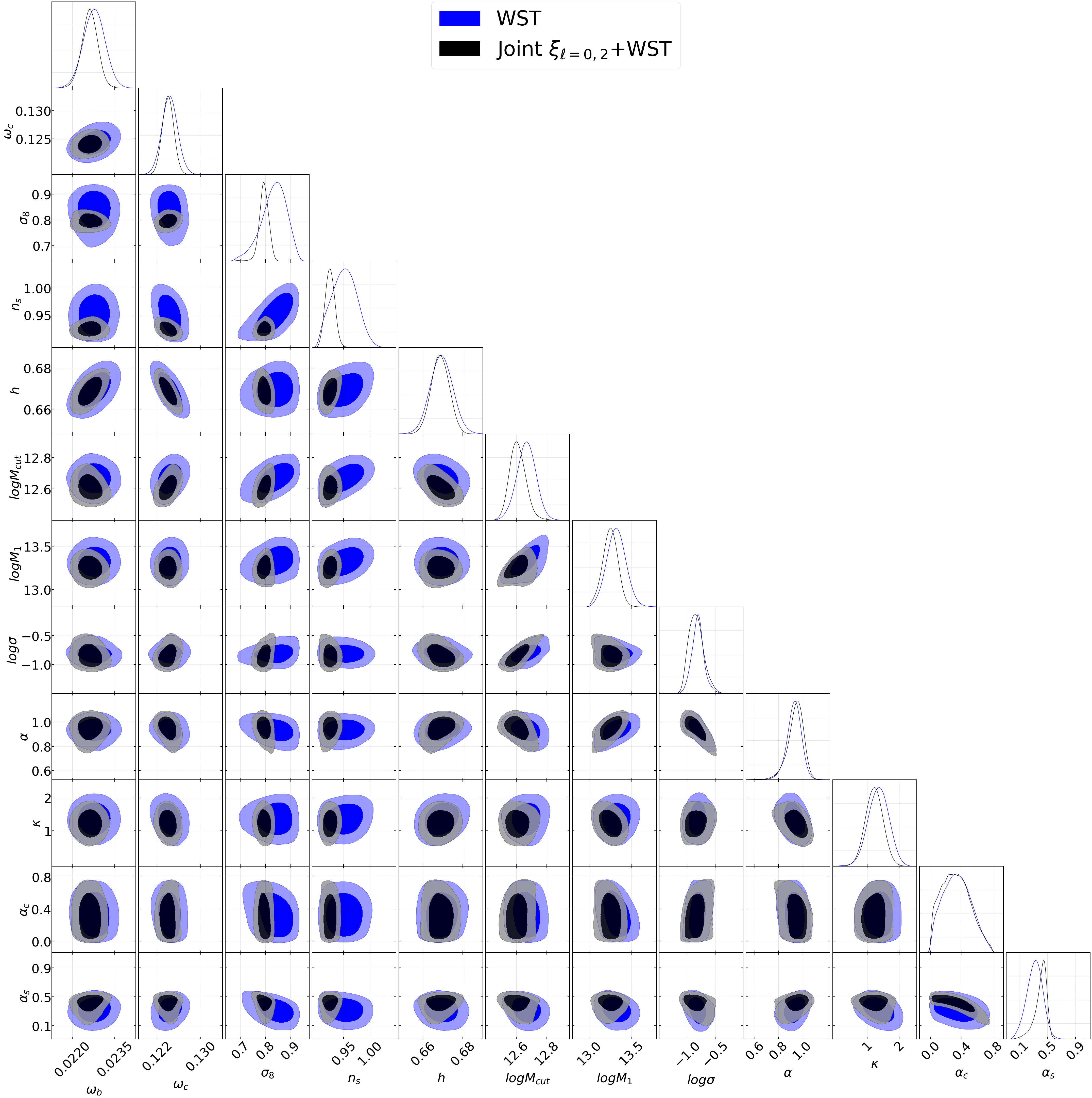}
\caption{Marginalized constraints on the full set of cosmological+HOD parameters obtained using the WST coefficients (blue) and the joint combination of WST+correlation function multipoles (black) in order to analyze the BOSS CMASS observations. The results shown above were obtained after imposing a BBN Gaussian prior on the value of $\omega_b = 0.02268 \pm 0.00038$.}
\label{Fig:Cornerallparams}
\end{figure*}

In this section, we present the constraints obtained on the full set of HOD parameters of our WST and joint WST + correlation function analyses, shown in Fig.~\ref{Fig:Cornerallparams} and Table~\ref{table:3}. We find that the WST alone is capable of constraining all HOD parameters of our model, with only modest additional improvements delivered after the inclusion of the 2-point correlation function. Our analysis hints at a preference for non-zero velocity biases both for the central and also for the satellite galaxies, through the corresponding inferred values for parameters $\alpha_c$ and $\alpha_s$. Even though the former result is in agreement with the small-scale CMASS reanalysis of Ref. \citep{2022MNRAS.515..871Y}, it is interesting that the same work did not find a preference for a satellite bias. We defer a more detailed investigation of this matter to future work, which will extend our analysis to equally small scales. 

{
\begin{table}
\centering

\setlength{\extrarowheight}{2.6pt}
\begin{tabular}{|p{1.1cm}||p{3.45cm}||p{3.55cm}|}
\hline 
& WST & Joint 2-point c.f.+WST\\
\end{tabular}

\centering
\begin{tabular}{|p{1.1cm}||p{1.1cm}|p{2.2cm}||p{1.2cm}|p{2.2cm}|}

\hline 
 & Best-fit& Mean$\pm \sigma$& Best-fit& Mean$\pm \sigma$\\
\hline 
$\log{\rm M}_{\rm cut}$ & $12.681$&$12.668^{+0.068}_{-0.068}$&$12.608$&$12.613^{+0.045}_{-0.060}$\\
\hline
$\log{\rm M}_{\rm 1}$ &13.34&$13.33^{+0.13}_{-0.13}$&$13.252$&$13.25^{+0.11}_{-0.11}$\\
\hline
$\log\sigma$ &$-0.783$&$-0.823^{+0.11}_{-0.097}$&$-0.829$&$-0.87^{+0.25}_{-0.25}$\\
\hline
$\alpha$ &$0.921$&$0.934^{+0.064}_{-0.054}$&$0.943$&$0.944^{+0.077}_{-0.049}$\\
\hline
$\kappa$ &$1.336$&$1.36^{+0.32}_{-0.32}$&$1.236$&$1.22^{+0.28}_{-0.28}$\\
\hline
$\alpha_\mathrm{c}$ &$0.322$&$0.34^{+0.17}_{-0.20}$&$0.367$&$0.32^{+0.16}_{-0.22}$\\
\hline
$\alpha_\mathrm{s}$ &$0.306$&$0.32^{+0.12}_{-0.11}$&$0.411$&$0.408^{+0.099}_{-0.049}$\\
\hline
\end{tabular}

\caption{Best-fit values, mean values and $68\%$ confidence intervals for the 7 nuissance HOD parameters of our base likelihood analysis using the WST coefficients (left) and the joint analysis of the 2-point correlation function + WST (right). The mean values are presented in the format `${\rm mean}^{+1 \sigma}_{-1 \sigma}$'.}
\label{table:3}
\end{table}
}

%%%%%%%%%%%%%%%%%%%%%%%%%%%%%%%%%%%%%%%%%%%%%%%%%%%%
\section{WST constraints without emulator error}\label{Appsec:Cemu}

\begin{figure}[ht!]
\centering 
\includegraphics[width=0.49\textwidth]{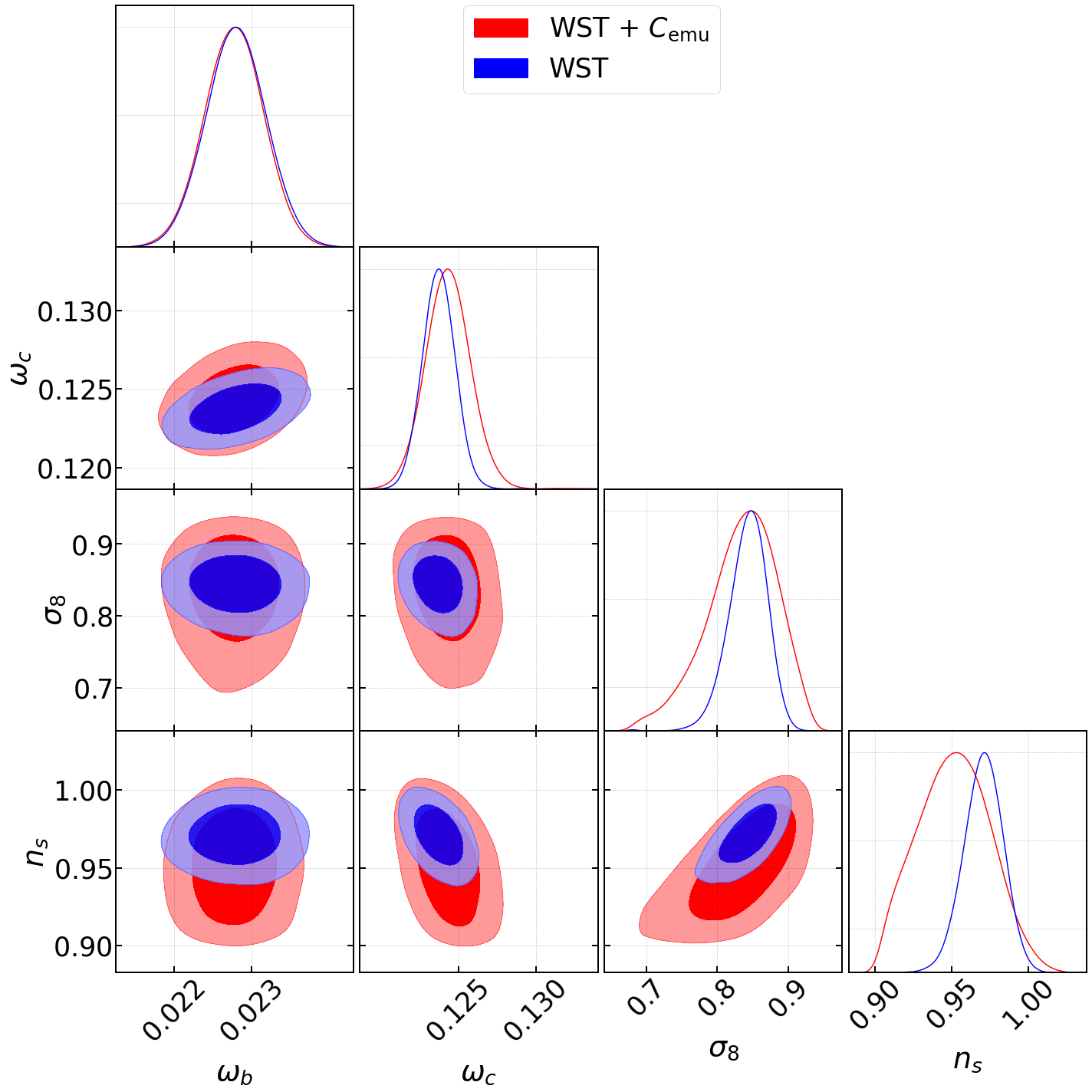}
\caption{Marginalized constraints on the $\Lambda$CDM cosmological parameters obtained using the WST coefficients without the inclusion of the emulator error, $C_{\rm emu}$, in Eq. \eqref{eq:covmatall}, shown in the blue contours. The result of the main WST analysis using the full covariance (originally shown in Fig.~\ref{Fig:Corner5params}) is also plotted in red, for comparison.}
\label{Fig:Cemucomp}
\end{figure}

In Section \S\ref{subsec:likelihood}, we explained how the residual emulator error, $C_{\rm emu}$, was treated as an additional covariance contribution that we added to the overall error budget, through Eq. \eqref{eq:covmatall}. In order to illustrate the impact of this factor to the WST constraints, and also to better facilitate the comparison with our previous work \citep{PhysRevD.106.103509} (which did not account for the emulator error), we repeat the WST analysis using the contribution from the \textsc{Patchy} mocks only (that is, the first term in Eq. \eqref{eq:covmatall}) and contrast it against the full result, in Fig.~\ref{Fig:Cemucomp}.
We notice that the inclusion of the emulator error $C_{\rm emu}$ leads to a substantial increase in the 1$\sigma$ errors for parameters $\sigma_8$ and $n_s$, in particular, with the corresponding impact being much less significant for $\omega_c$. When we neglect this term, on the other hand, and as we also pointed out in the main text, the constraints become much tighter and comparable to the ones of our previous analysis \citep{PhysRevD.106.103509} in the case of $\sigma_8$ and $n_s$. We stress that this result should be interpreted with caution, given that the emulator error has not been accounted for. It does serve, nevertheless, as an indication of the intrinsic constraining power of the WST in the limit of zero emulation error. In order for this potential to actually be exploited by the next stage of precise spectroscopic observations, however, higher accuracy emulators and more precise characterization of the emulator error will be necessary. Whether and how these goals can be achieved is a matter of intense study.

%%%%%%%%%%%%%%%%%%%%%%%%%%%%%%%%%%%%%%%%%%%%%%%%%%%%
\section{Impact of priors on $\omega_b$}\label{Appsec:omegabpriors}

\begin{figure}[ht!]
\centering 
\includegraphics[width=0.49\textwidth]{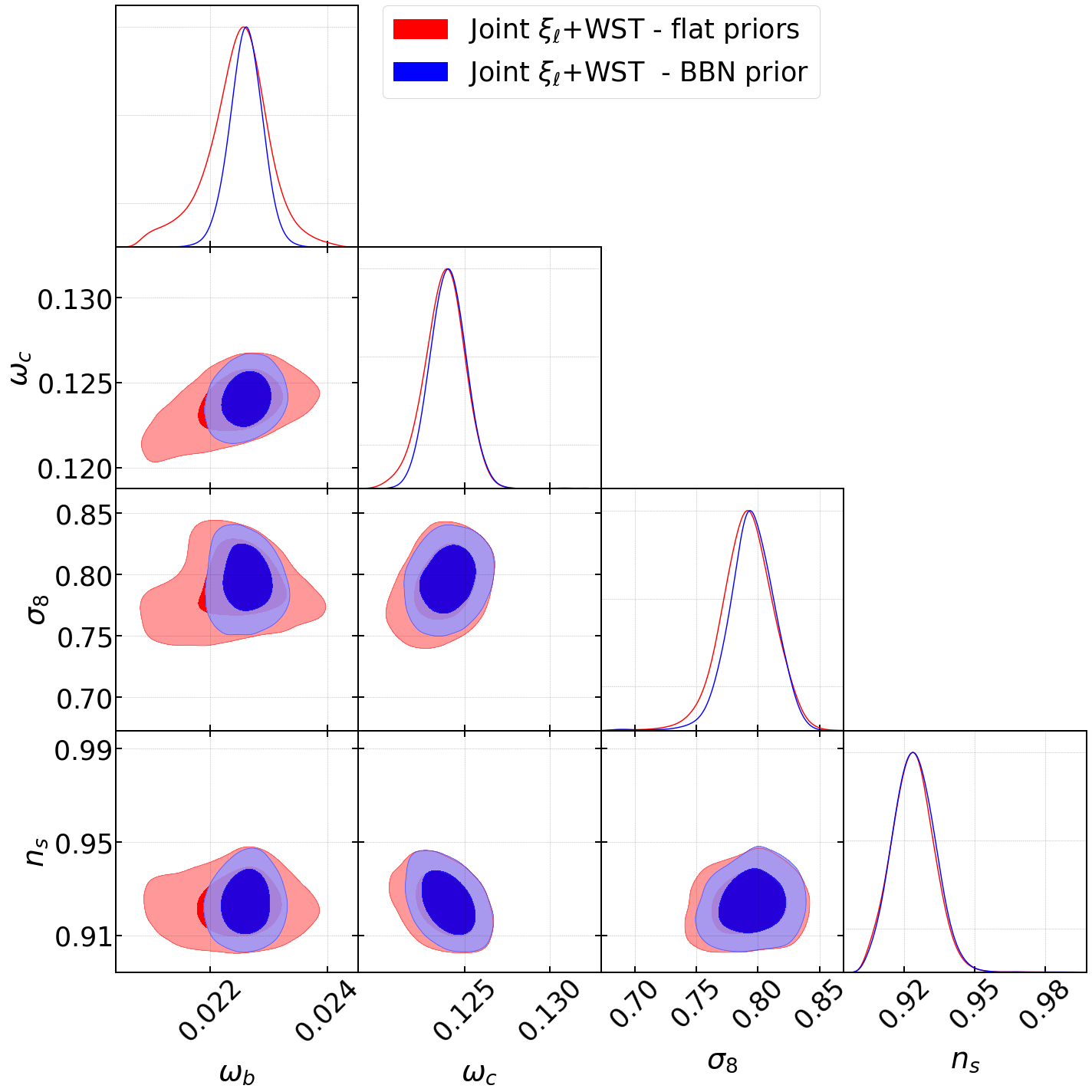}
\caption{Marginalized constraints on the $\Lambda$CDM cosmological parameters obtained from the joint WST + correlation function analysis using a flat prior on the value of $\omega_b$ (red), as opposed to the main analysis that used a Gaussian prior \eqref{omegabprior}, in blue (and originally presented in Fig.~\ref{Fig:Corner5params}).}
\label{Fig:BBNcomp}
\end{figure}

Our main analysis used a tight BBN prior on the value of $\omega_b$, from Eq. \eqref{omegabprior}. In this appendix we repeat our joint WST + correlation function analysis using a flat $\omega_b$ prior, instead, and demonstrate the comparison between the two results in Fig.~\ref{Fig:BBNcomp}. Remarkably, we find that the joint analysis is also able to accurately constrain $\omega_b$, as well as the rest of the parameters, using completely uninformative priors. The corresponding increase in the 1$\sigma$ errors is 90$\%$ for $\omega_b$ and no more than 10$\%$ for the rest of the 3 parameters.

\section{Sensitivity to small scales}\label{Appsec:kmax}

\begin{figure}[ht!]
\centering 
\includegraphics[width=0.49\textwidth]{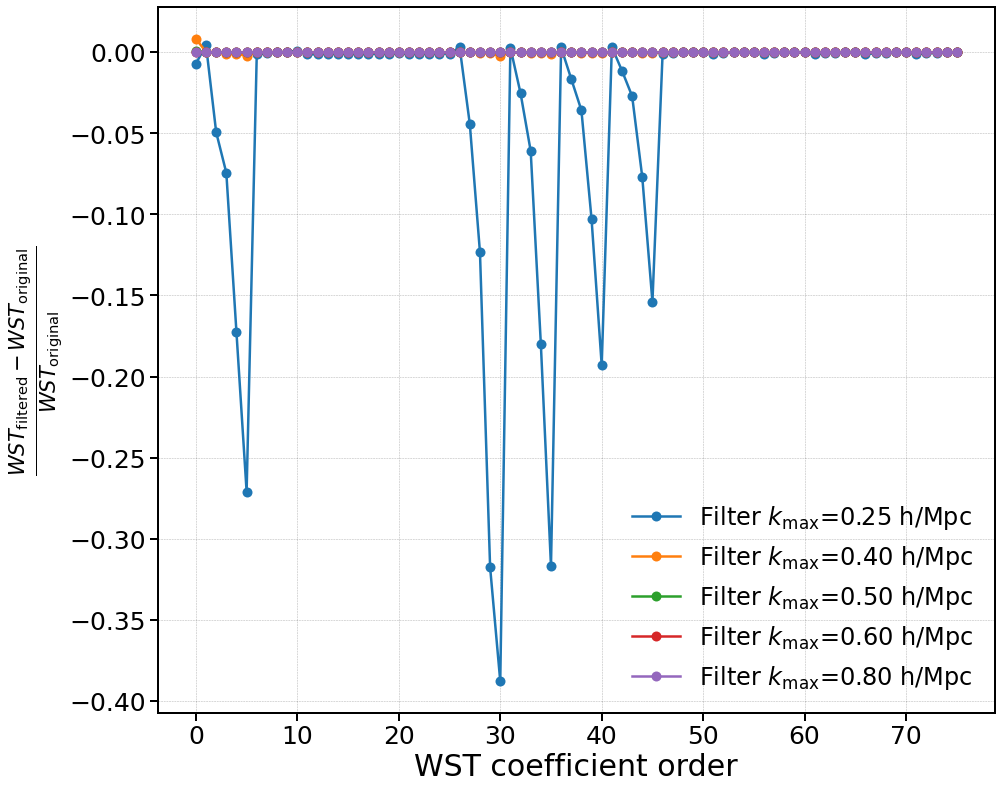}
\caption{Fractional changes to the WST data vector when a sharp top-hat filter with various $k_{\rm max}$ cut-off values is applied to the galaxy field before the evaluation in Eq. \eqref{eq:WSTcoeff:sol}, with respect to the original evaluation of our main analysis. This example evaluation corresponds to the NGC part of the BOSS dataset.}
\label{Fig:ktests}
\end{figure}

In principle, the solid harmonic wavelets that we use in this analysis do not have a finite support neither in real or Fourier space. The Fourier transform of the radial part of Eq. \eqref{solid:sup} is a Gaussian for $\ell=0$, that can extend to higher $k$ for  $\ell > 0$ values. Even though in practice this can be controlled through a sufficiently conservative combination of the Gaussian width and grid size, as we did in this application, we need to explicitly make sure that our WST analysis does not extract information from smaller scales than originally intended. We confirm this fact through the following test: we first apply a sharp top-hat filter in k-space to our galaxy field and, after going to real space, use this filtered field instead as the input into the WST scattering network \eqref{eq:WSTcoeff:sol}. This addition imposes a sharp k-space cut off, which would remove any potential undesired contributions from higher frequencies (smaller scales). In Fig.~\ref{Fig:ktests}, we plot the fractional change to the WST data vector obtained from the BOSS NGC data when this filtering is applied for various cut-offs, compared to the original prediction using just the Gaussian-like smoothing from Eq. \eqref{solid:sup}, with $\sigma=0.80$ and $N_{\rm grid}=270$. Our 2-point correlation function benchmark analysis includes scales down to 8 Mpc/h, corresponding to a $k_{max} = 0.8$ h/Mpc in the Fourier space. As we see in Fig.~\ref{Fig:ktests}, imposing this sharp cut-off leads to no measurable changes in the WST data vector compared to the original one, indicating no sensitivity to $k > 0.8$ h/Mpc. Imposing progressively stricter cutoff values leads to growing differences in the data vector, as we remove scales that our wavelets were originally sensitive to. If we restrict our focus on  wavemodes $k \leq 0.25$ h/Mpc, the changes in the data vector are the most pronounced, as expected, since that would discard the majority of the nonlinear information contained in the galaxy field. Different combinations of the Gaussian width and/or grid size lead to a different spatial support, which can similarly be further contained with the sharp k-space filter. We also confirmed that the behavior in Fig.~\ref{Fig:ktests} holds not just for the BOSS data, but also for our simulation-based model predictions across the prior space. These findings confirm that our specific choices of Gaussian width, grid size and harmonic order for the wavelet analysis were conservative enough and did not access scales smaller than the ones of the correlation function benchmark analysis. 

We also note that wavelets which are explicitly designed to have a finite support in Fourier space, such as e.g. the ones used in \citep{2022arXiv220407646E,2023arXiv231015250R}, are a natural next improvement to the above approach, that we are actively working on implementing in advance of the application to the next generation of spectroscopic data.

\clearpage

\end{document}